\def\year{2020}\relax
\documentclass[letterpaper]{article} 
\usepackage{aaai20}  
\usepackage{times}  
\usepackage{helvet} 
\usepackage{courier}  
\usepackage[hyphens]{url}  
\usepackage{graphicx} 
\usepackage{graphicx}  
\title{Distributed Machine Learning through Heterogeneous Edge Systems}

\author{
Hanpeng Hu,\textsuperscript{\rm 1}
Dan Wang,\textsuperscript{\rm 2} 
Chuan Wu\textsuperscript{\rm 1} \\
\textsuperscript{\rm 1}The University of Hong Kong, \textsuperscript{\rm 2}The Hong Kong Polytechnic University \\
hphu@cs.hku.hk, csdwang@comp.polyu.edu.hk, cwu@cs.hku.hk.}

\RequirePackage{totpages}
\usepackage{fancyhdr}
\usepackage{booktabs}
\usepackage{amsmath} 
\usepackage{amssymb} 
\usepackage{amsthm}
\newtheorem{theorem}{Theorem} 
\usepackage{graphicx} 

\usepackage{algorithm}
\usepackage{algorithmicx}
\usepackage{algpseudocode}

\usepackage{enumitem}
\usepackage{url}

\usepackage[title]{appendix}

\newcommand{\ignore}[1]{{}}

\begin{document}

\maketitle

\begin{abstract}
Many emerging AI applications request distributed machine learning (ML) among edge systems (e.g., IoT devices and PCs at the edge of the Internet), where data cannot be uploaded to a central venue for model training, due to their large volumes and/or security/privacy concerns.
Edge devices are intrinsically {\em heterogeneous} in computing capacity, posing significant challenges to parameter synchronization for parallel training with the parameter server (PS) architecture.
This paper proposes {\em ADSP}, a parameter synchronization model for distributed machine learning (ML) with heterogeneous edge systems.
Eliminating the significant waiting time occurring with existing parameter synchronization models, the core idea of {\em ADSP} is to let faster edge devices continue training, while committing their model updates at strategically decided intervals. We design algorithms that decide time points for each worker to commit its model update, and ensure not only global model convergence but also faster convergence. Our testbed implementation and experiments show that {\em ADSP} outperforms existing parameter synchronization models significantly in terms of ML model convergence time, scalability and adaptability to large heterogeneity.
\end{abstract} 

\section{Introduction}
Many edge-based AI applications have emerged in recent years, where various edge systems (e.g., PCs, smart phones, IoT devices) collect local data, collaboratively train a ML model, and use the model for AI-driven services. For example, smart cameras are deployed in surveillance systems \cite{ai_solution_edge,park2018wireless}, which capture local images/videos and train a global face recognition model aggregately. In Industry AI Operations (AIOps) \cite{qu2017next}, chillers in a building or an area collect temperature and electricity consumption data in the households, and derive a global COP (Coefficient of Performance) prediction model \cite{chen2019data}.

A straightforward way of training a global model with data collected from multiple edge systems is to send all data to a central venue, e.g., a cloud data center, and train the datasets using a ML framework, such as TensorFlow \cite{2016abadi-tensorflow}, MXNet \cite{2015chen-mxnet} and Caffe2 \cite{hazelwood2018applied}). Such a `data aggregation $\rightarrow$ training' approach may well incur large network bandwidth cost, due to large data transmission volumes and continuous generation nature of the data flow, as well as data security and privacy concerns. To alleviate these issues, collaborative, distributed training among edge systems has been advocated \cite{tao2018esgd}, where each edge system locally trains the dataset it collects, and exchanges model parameter {\em updates} (i.e., gradients) with each other through parameter servers \cite{wang2018adaptive_federated,konevcny2016federated,park2018wireless} (a.k.a., geo-distributed data parallel training). 

Edge systems are intrinsically heterogeneous: their hardware configurations can be vastly different, leading to different computation and communication capacities. This brings significant new issues on parameter synchronization among the edge workers. In a data center environment, synchronous training (i.e., Bulk Synchronous Parallel (BSP) \cite{2013ssp} \cite{li2013parameter} \cite{low2012distributed}) is adopted by the majority of production ML jobs (based on our exchanges with large AI cloud operators), given the largely homogeneous worker configuration: each worker trains a mini-batch of input data and commits computed gradients to the PS; the PS updates global model after receiving commits from all workers, and then dispatches updated model parameters to all workers, before each worker can continue training its next mini-batch. In the edge environment, the vastly different training speeds among edge devices call for a more asynchronous parameter synchronization model, to expedite ML model convergence.

{\em Stale Synchronous Parallel (SSP)} \cite{2013ssp} and {\em Totally Asynchronous Parallel (TAP)} \cite{2017hsieh-gaia} are representative asynchronous synchronization models. With TAP, the PS updates the global model upon commit from each individual worker, and dispatches updated model immediately to the respective worker; it has been proven that such complete asynchrony cannot ensure model convergence \cite{2017hsieh-gaia}. SSP enforces bounded asynchronization:  fast workers wait for slow workers for a bounded difference in their training progress, in order to ensure model convergence. A few recent approaches have been proposed to further improve convergence speed of asynchronous training\cite{hadjis2016omnivore,wang2018adaptive} (see more in Sec.~\ref{sec:relatedwork}). 

We investigate how existing parameter synchronization models work in a heterogeneous edge environment with testbed experiments (Sec.~\ref{motivation}), and show that the waiting time (overall model training time minus gradient computation time) is still more than 50\% of the total training time with the representative synchronization models. 

Aiming at minimizing the waiting time and optimizing computing resource utilization, we propose {\em ADSP} (ADaptive Synchronous Parallel), a new parameter synchronization model for distributed ML with heterogeneous edge  systems. Our core idea is to let faster workers continue with their mini-batch training all the time, while enabling all workers to commit their model updates at the same strategically decided intervals, to ensure not only model convergence but also faster convergence. The highlights of {\em ADSP} are summarized as follows:

$\triangleright$ {\em ADAP} is tailored for distributed training in heterogeneous edge systems, which fully exploits individual workers' processing capacities by eliminating the waiting time.

$\triangleright$ {\em ADSP} actively controls the parameter update rate from each worker to the PS, to ensure that the total number of commits from each worker to the PS is roughly the same over time, no matter how fast or slow each worker performs local model training. Our algorithm exploits a momentum-based online search approach to identify the best cumulative commit number across all workers, and computes the commit rates of individual workers accordingly. ADSP is proven to converge after a sufficient number of training iterations.

$\triangleright$ We have done a full-fledged implementation of {\em ADSP} and evaluated it with real-world edge ML applications. Evaluation results show that it outperforms representative parameter synchronization models significantly in terms of model convergence time, scalability and adaptability to large heterogeneity.

\section{Background and Motivation}
\label{sec:relatedwork}

\subsection{SGD in PS Architecture}

Stochastic Gradient Descent (SGD) is the widely used algorithm for training neural networks \cite{hadjis2016omnivore,2016abadi-tensorflow}. Let $W_t$ be the set of global parameters of the ML model at $t$. A common model update method with SGD is:

\begin{equation}\label{eq:sgdMomentum}
\begin{aligned}
W_{t+1} = W_t  - \eta \nabla \ell(W_t)+ \mu (W_{t} - W_{t-1})
\end{aligned}
\end{equation}

\noindent where $\nabla \ell(W_t)$ is the gradient, $\eta$ is the learning rate, and $\mu \in [0, 1]$ is the {\em momentum} introduced to accelerate the training process, since it accumulates gradients in the right direction to the optimal point \cite{polyak1964some,sutskever2013importance}. 

In widely adopted data-parallel training with the parameter server (PS) architecture\cite{2014chilimbi-Adam}, SGD update rule can be applied at both the workers and the PS \cite{jiang2017heterogeneity}. Each worker holds a local copy of the ML model, its local dataset is divided into mini-batches, and the worker trains its model in an iterative fashion: in each step, the worker calculates gradients of model parameters using one mini-batch of its data, and may commit its gradients to the PS and pull the newest global model parameters from the PS. The PS updates the global model using Eqn.~(\ref{eq:sgdMomentum}) with gradients received from the workers and a \textit{global learning rate} $\eta$. In the case that a worker does not synchronize model parameters with the PS per step, the worker may carry out local model updates using computed gradients according to Eqn.~(\ref{eq:sgdMomentum}), where the gradients are multiplied by a \textit{local learning rate} $\eta^{\prime}$.

\subsection{Existing Parameter Synchronization Models} 
A parameter synchronization model specifies when each worker commits its gradients to the PS and whether it should be synchronized with updates from other workers; it critically affects convergence speed of model training. Three representative synchronization models, BSP, SSP and TAP, have been compared in \cite{2017hsieh-gaia}, which proves that BSP and SSP guarantee model convergence whereas TAP does not. Training convergence with BSP is significantly slower than SSP \cite{2013ssp}, due to BSP's strict synchronization barriers. Based on the three synchronization models, many studies have followed, aiming to reduce the convergence time by reducing communication contention or overhead  \cite{chenround,lin2017deep,sun2016timed}, adjusting the learning rate \cite{jiang2017heterogeneity}, and others  \cite{zhang2018stay}.
ADACOMM \cite{wang2018adaptive} allows accumulating local updates before committing to the PS, and adopts BSP-style synchronization model, i.e., all workers run $\tau$ training steps before synchronizing with the PS. It also suggests reducing the commit rate periodically according to model loss; however, the instability in loss values during training and the rapidly declining commit rate are not ideal for expediting training  (according to our experiments).  

Aiming at minimizing waiting time among heterogeneous workers, our synchronization model, ADSP, employs an online search algorithm to automatically find the optimal/near-optimal update commit rate for each worker to adopt.

\subsection{Impact of Waiting} \label{motivation}
\begin{figure}[!th]
\begin{center}
  \includegraphics[width = .95\columnwidth]{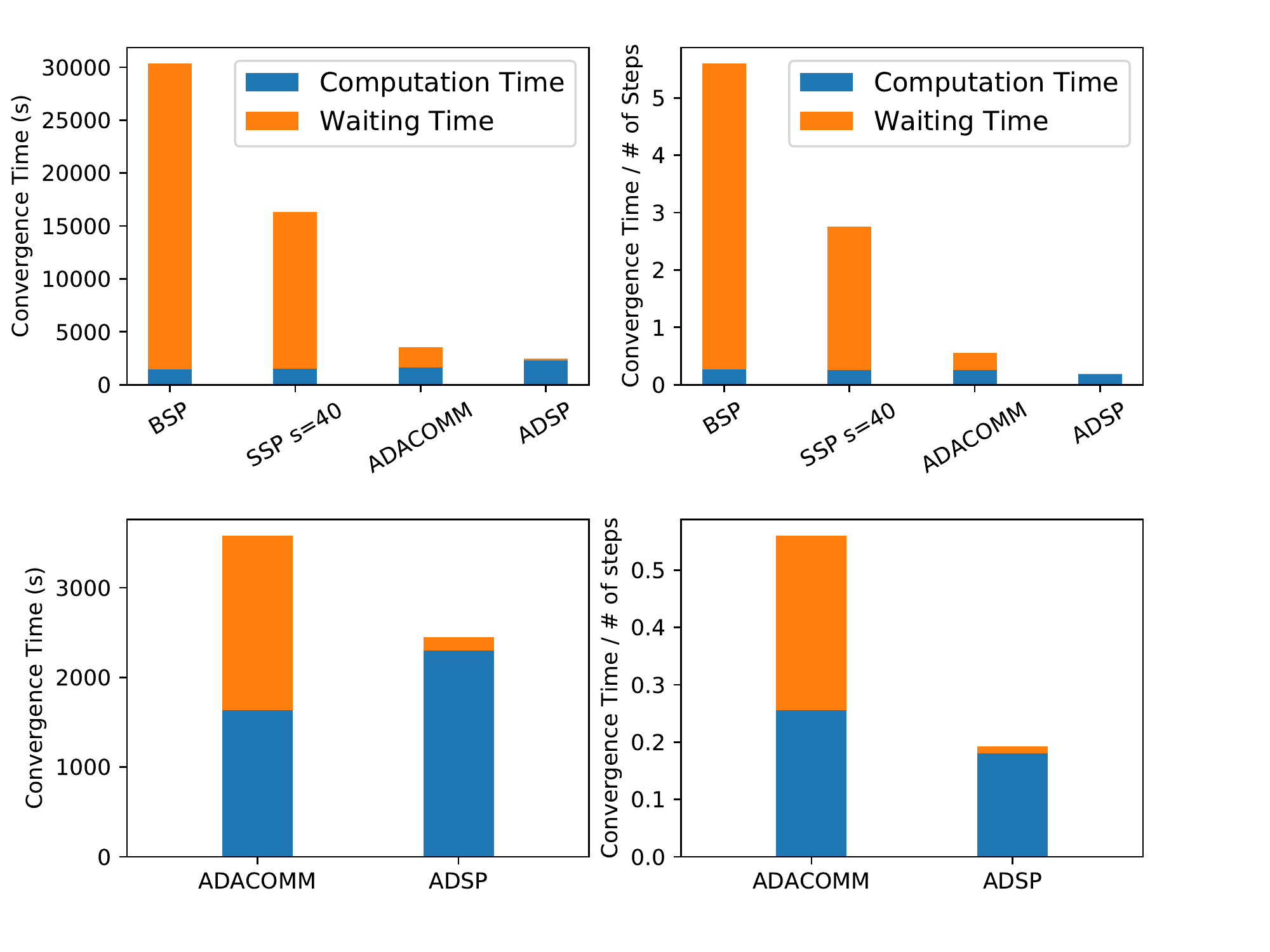}
  \caption{Training time breakdown with different parameter synchronization models.}
  \label{figure:waiting_time}
\end{center}
\end{figure}

We divide the time a worker spends in each training step into two parts: (i) the {\em computation time}, to carry out backward propagation to compute gradients/apply model updates and forward propagation to produce output \cite{2015chen-mxnet}; and (ii) the \textit{waiting time}, including the time for exchanging gradients/parameters with the PS and the blocked time due to synchronization barrier (i.e., the time when the worker is not doing computation nor communication). 

We experiment with representative synchronization models to investigate their waiting time incurred. We train a convolutional neural network (CNN) model on the Cifar10 dataset \cite{cifar10} with 1 PS and 3 workers with heterogeneous computation capacities (time ratio to train one mini-batch is 1:1:3). Fig.~\ref{figure:waiting_time} shows the convergence time (overall training time to model convergence) and the average time spent per training step, incurred with BSP, SSP, and ADACOMM (See Sec.~\ref{sec:evaluation} for their details). TAP is not compared as it has no convergence guarantee. The computation/waiting time is averaged among all workers. We see that with heterogeneous workers, the waiting time dominates the overall training time with BSP and SSP, and their overall convergence time and time spent per training step are long. With ADACOMM, the waiting time and overall training time are much shorter. Nevertheless, its waiting time is still close to half of the total training time, i.e., the effective time used for model training is only around 50\%, due to its relative conservative approach on local model updates. 

Our key question is: what is the limit that we can further reduce the waiting time to, such that time is spent most efficiently on model computation and convergence can be achieved in the most expedited fashion? Our answer, ADSP, allows fast workers to keep training while maintaining approximately the same gradient commit rates among all workers. Fig.~\ref{figure:waiting_time} shows that the waiting time is minimized to a negligible level with ADSP, as compared to the computation time. As such, almost all training time is effectively used for model computation and fast model convergence is achieved.

\section{ADSP Overview}\label{section:overview}
We consider a set of heterogeneous edge systems and a parameter server (PS) located in a datacenter, which together carry out SGD-based distributed training to learning a ML model. ADSP (ADaptive Synchronous Parallel) is a new parameter synchronization model for this distributed ML system. The design of ADSP targets the following goals: 
(i) make full use of the computation capacity of each worker; 
(ii) choose a proper commit rate to balance the tradeoff between hardware efficiency (utilization of worker computing capacity) and statistical efficiency (i.e., reduction of loss per training step), in order to minimize the overall time taken to achieve model convergence; 
(iii) ensure model convergence under various training speeds and bandwidth situations at different workers.

\begin{figure}[!t]
\begin{center}
  \includegraphics[width = .95\columnwidth]{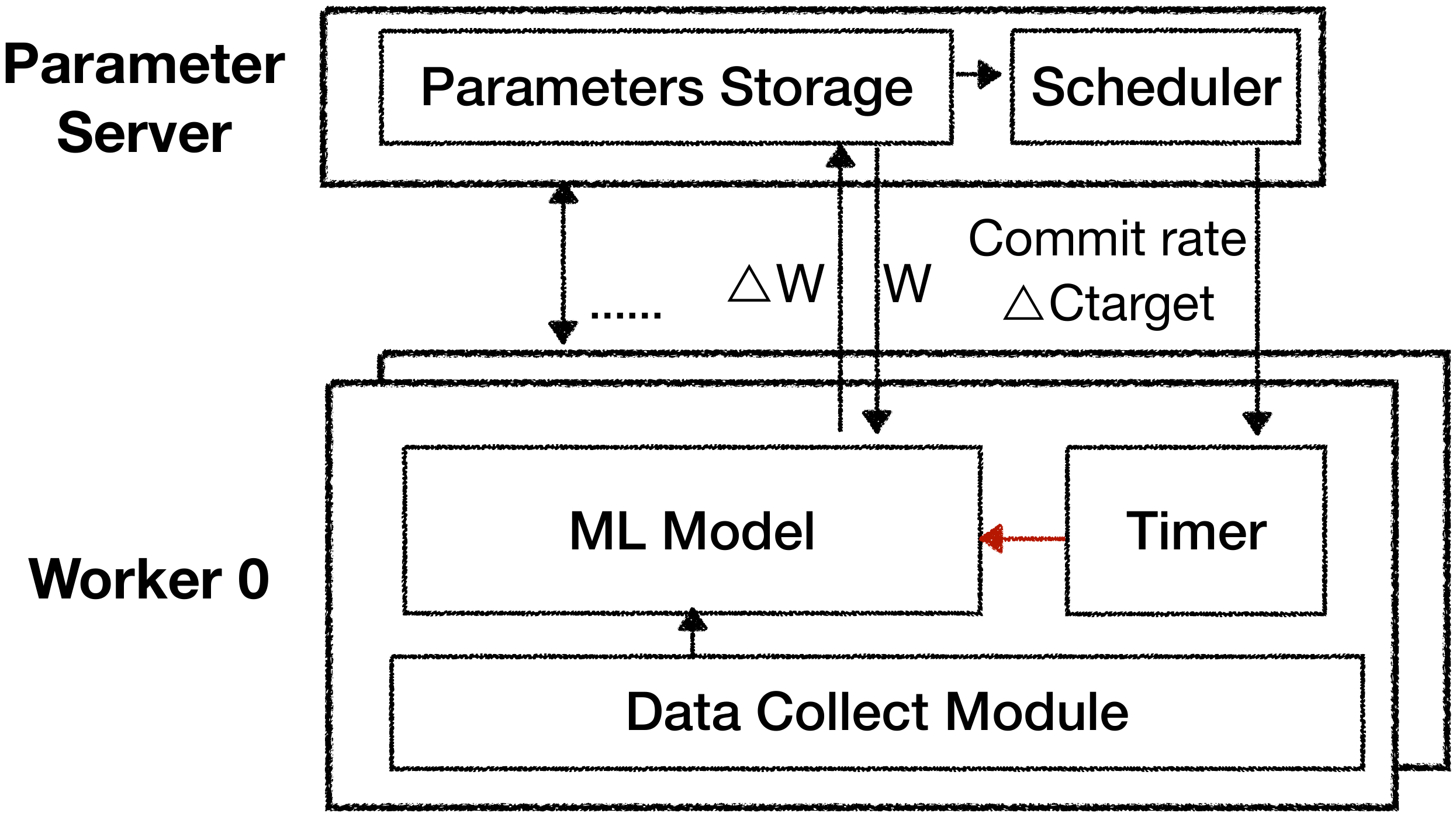}
  \caption{ADSP workflow.}
  \label{figure:STrain_arc}
\end{center}
\end{figure}

With ADSP, time is divided into equal-sized slots of duration $\Gamma>0$: $0, \Gamma, 2\Gamma, \ldots$, which we call as \textit{check periods}. We refer to time points $\Gamma, 2\Gamma, \ldots, p\Gamma, \ldots$, as {\em checkpoints}. More precisely, we define the process of a worker sending computed gradients to the PS as a \textit{commit}, and the number of commits from worker $i$ during a check period as \textit{commit rate} $\Delta C_{target}^{i}$. ADSP consists of two modules: 1) a novel synchronization model, which allows faster edge systems to perform more training before each update to the PS, and ensures that the commit rates of all worker are the same; 2) a global commit rate search algorithm, which selects an appropriate commit rate for all workers to pursue, in order to achieve fast convergence. 
 
 Let $c_i$ denote the total number of commits from worker $i$ to the PS, since the very beginning. At each checkpoint, we compute the target total number of commits that each worker is expected to have submitted by the next checkpoint, $C_{target}$, and adjust the commit rate of each worker $i$ in the next check period as $\Delta C_{target}^{i} = C_{target} - c_i$, respectively.

Fig.~\ref{figure:STrain_arc} shows the workflow of our ADSP model. The data produced/collected at each edge system/worker is stored into training datasets. For each mini-batch in its dataset, an edge system computes a \textit{local update} of model parameters, i.e., gradients, using examples in this mini-batch. After training one mini-batch, it moves on to train the next mini-batch and derives another local update. Worker $i$ pushes its \textit{accumulative update} (i.e., sum of all gradients it has produced since last commit multiplied with the local learning rate) according to the commit rate $\Delta C_{target}^{i}$. A scheduler adjusts and informs each end system of the target commit rate $\Delta C_{target}^i$ over time. Upon receiving a commit from worker $i$, the PS multiplies the accumulated update with the \textit{global learning rate} \cite{jiang2017heterogeneity} and then updates the global model with it; worker $i$ then pulls updated parameters from the PS and continues training the next mini-batch.

\section{ADSP Algorithms and Analysis}

It is common to have large heterogeneity among edge systems, including different computation power and network delays to the datacenter hosting the PS. Our core idea in designing ADSP is to {\em adapt} to the heterogeneity, i.e., to transform the training in heterogeneous settings into homogeneous settings using a {\em no-waiting} strategy: \textit{we allow different workers to process different numbers of mini-batches between two commits according to their training speed, while ensuring the number of commits of all the workers approximately equal at periodical checkpoints.} To achieve this, we mainly control the hyper-parameter, {\em commit rate}, making faster workers accumulate more local updates before committing their updates, so as to eliminate the waiting time. By enforcing approximately equal numbers of commits from all the workers over time, we can ensure model convergence.

\subsection{The Impact of $C_{target}$ on Convergence}  

The target total number of commits to be achieved by each worker by the next checkpoint, $C_{target}$, decides commit rate of each worker $i$ within the next check period, as $\Delta C_{target}^i = C_{target} - c_i$ ($c_i$ is $i$'s current total commit number). The commit rate has a significant impact on the training progress: if $\Delta C_{target}^i$ is large, a slow worker may fail to achieve that many commits in the next period, due to the limited compute capacity; even if it can hit the target, too many commits may incur high communication overhead, which in turn slows down the training process. On the other hand, if the number of target commits is too small, which implies that each end system commits its gradients after many steps of mini-batch training using local parameters, large difference and significant staleness exist among the model copies at different workers, which may adversely influence model convergence as well. 

\begin{figure}
\begin{center}
  \includegraphics[width = .95\columnwidth]{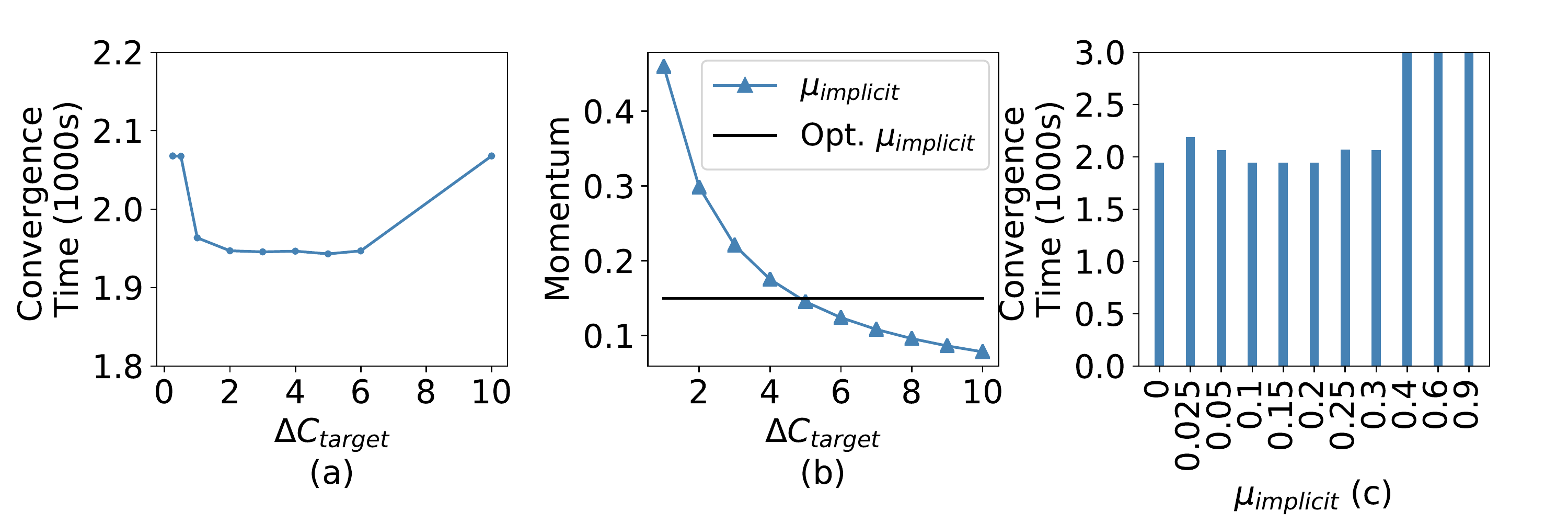}
  \caption{(a) the impact of $C_{target}$ on convergence time; (b) an illustration of $\mu_{implicit}$ ; (c) convergence time with different $\mu_{implicit}$ values.}   
  \label{figure:vary_commit_rate_mu}
\end{center}
\end{figure}

To illustrate this, we train a CNN model on the Cifar10 dataset \cite{cifar10} with 1 PS and 3 workers (time ratio to train one mini-batch is 1:1:3), where all workers keep training their mini-batches and commit gradients to the PS at the same commit rate $\Delta C_{target}$ over time. We vary the value of $\Delta C_{target}$ in different runs of the experiment. Fig.~\ref{figure:vary_commit_rate_mu}(a) shows that with the increase of $\Delta C_{target}$, the model convergence time becomes smaller at first and then increases. This is consistent with our discussions above.

We next quantify the effect of the commit rate $\Delta C_{target}$ on model convergence. Suppose that all $m$ workers communicate with the PS independently. Let $U(W_t)$ denote the accumulative local updates that a worker commits when the global model is $W_t$, and $v_i$ denote the number of steps that worker $i$ can train per unit time. We have the following theorem. 

\begin{theorem}\label{local_model}
Set the momentum $\mu$ in the SGD update formula (\ref{eq:sgdMomentum}) to zero. The expected SGD update on the global model is equivalent to 
\begin{eqnarray}
\mathbb{E}(W_{t+1}\! - W_{t})\!  = \!(1-p)\mathbb{E}(W_{t}\! - W_{t-1}) - p \eta \mathbb{E} U(W_t) \label{eq:theorem1} \\
\mbox{where } ~ p = 1/(1 + (1 - 1/m)\sum_{i=1}^{m}\frac{\Gamma}{\Delta C_{target}^i v_i}) \label{eq:theorem1_p}
\end{eqnarray}
\end{theorem}

\noindent The detailed proof is given in the supplemental file. Compared to the SGD update formula in Eqn.~(\ref{eq:sgdMomentum}), the result is interesting: with our ADSP model, staleness induced by cumulative local updates can be considered as inducing an extra momentum term (i.e., $1-p$) into the SGD update equation. To distinguish this term from the original momentum $\mu$ in Eqn.~(\ref{eq:sgdMomentum}), we refer to this term as the \textit{implicit momentum}, denoted by $\mu_{implicit}=1-p$. As we increase $\Delta C_{target}^i$, the implicit momentum becomes smaller according to Eqn.~(\ref{eq:theorem1}). 

With the same CNN training experiments as above, Fig.~\ref{figure:vary_commit_rate_mu}(b) illustrates how $1-p$ varies with $\Delta C_{target}$ (according to Eqn.~(\ref{eq:theorem1_p})). The optimal momentum is derived based on  Fig.~\ref{figure:vary_commit_rate_mu}(c), where we vary the value of $\mu_{implicit}$ in Eqn.~(\ref{eq:theorem1}) in our experiments, and show how the time taken for model convergence varies with different $\mu_{implicit}$ values. Inspired by the observations, we seek to identify the best commit rate $\Delta C_{target}$ for the workers, that decides the best $\mu_{implicit}$ to achieve the shortest convergence time.

\subsection{The Commit Rate Search Algorithm} 
\label{section:search}

We propose a local search method to identify a near-optimal commit rate to achieve the fastest convergence, exploiting the observations that the staleness induced by local updates can be converted to an implicit momentum term in SGD update and the implicit momentum decreases as we increase the commit rate. The algorithm is given in Alg.~\ref{algorithm:scheduler}, which is executed by the scheduler (Fig.~\ref{figure:STrain_arc}). 

In the algorithm, an {\em epoch} is a time interval containing multiple check periods, for commit rate adjustment. At the beginning of each \textit{epoch} (e.g., 1 hour), the scheduler performs the search for the optimal commit rates of workers in this epoch. We start with a small target total commit number $C_{target}$, allowing each worker to commit at least once in each check period; in this case, the commit rates $\Delta C_{target}^i$'s are small, asynchrony-induced implicit momentum is large, and the corresponding point in Fig.~\ref{figure:vary_commit_rate_mu}(b) is located to the left of the optimal momentum. Then the scheduler evaluates the training performance (i.e., loss decrease speed, to be detailed in Sec.~\ref{section:onlineSearch}) induced by $C_{target}$ and $C_{target}+1$, by running the system using commit rates computed based on the two values for a specific period of time (e.g., 1 minute). If $C_{target}+1$ leads to better performance, the scheduler repeats the search, comparing performance achieved by $C_{target}+1$ and $C_{target}+2$ further; otherwise, the search stops and the commit rates $\Delta C_{target}^i$'s decided by the current $C_{target}$ are used for the rest of this epoch. The rationale behind is that the optimal $C_{target}$ for each epoch is larger than the initial value ($\max_{i=1,\ldots,M}{c_i+1}$), so we only need to determine whether to increase it or not. 

\renewcommand{\algorithmicrequire}{\textbf{Scheduler:}}
\renewcommand{\algorithmicensure}{\textbf{Workers:}}
\begin{algorithm}[!th]
\caption{Commit Rate Adjustment at the Scheduler}
\label{algorithm:scheduler}
\begin{algorithmic}[1]
    \Function {\textsc{MainFunction}}{}
        \For {epoch e = 1, 2, \ldots}
			\State $C_{target}=\max_{i=1,\ldots,M}{c_i+1}$
            \State $C_{target} \leftarrow $ \textsc{DecideCommitRate}($C_{target}$)
            \State run \textsc{ParameterServer} and \textsc{Wokers} for the remaining time
        \EndFor 
    \EndFunction
    \Function {\textsc{DecideCommitRate}}{$C_{target}$}
            \State $r_1 \leftarrow$ \textsc{OnlineEvaluate}($C_{target}$)
            \State $r_2 \leftarrow$ \textsc{OnlineEvaluate}($C_{target} + 1$)
            \If {$r_2 > r_1$}
               \State \Return \textsc{DecideCommitRate}($C_{target} + 1$).
            \Else
               \State \Return $C_{target}$
            \EndIf
    \EndFunction
    \Function {\textsc{OnlineEvaluate}}{$C_{target}$}
        \For {$i$ = 0, 1, 2, \ldots,M}
            \State $\Delta C_{target}^i = C_{target} - c_i$
            \State Send $\Delta C_{target}^i$ to worker $i$
        \EndFor
        \State Training for 1 minute
        \State \Return reward $r$
    \EndFunction
\end{algorithmic}
\end{algorithm}

\subsubsection{Online Search and Reward Design. } \label{section:onlineSearch}
Traditional search methods are usually offline \cite{hadjis2016omnivore}, blocking the whole system when trying out a specific set of variable values and trying each configuration starting with the same system state. With an offline search method, one can select the best configuration by comparing the final loss achieved after running different configurations for the same length of time. However, such a search process incurs significant extra delay into the training progress and hence significant slowdown of model convergence. In Alg.~\ref{algorithm:scheduler}, we instead adopt an {\em online search} method (in DECIDECOMMITRATE()): we consecutively run each configuration for a specific time (e.g., 1 minute) without blocking the training process. 

To compare the performance of the configurations when they do not start with the same system state, we define a \textit{reward} as follows. The loss convergence curve of SGD training usually follows the form of $O(1/t)$ \cite{peng2018optimus}. We collect a few (time $t$, loss $\ell$) pairs when the system is running with a particular configuration, e.g., at the start, middle and end of the 1 minute period, and use them to fit the following formula on the left:
$$
\ell = \frac{1}{a_1^2 t + a_2} + a_3 \quad \Rightarrow \quad r = \frac{a_1^2}{\frac{1}{\ell - a_3} - a_2}
$$

\noindent where $a_1, a_2, a_3$ are parameters. Then we obtain the reward $r$ as the loss decrease speed, by setting $\ell$ to a constant and calculating the reciprocal of corresponding $t$. The target of the online search algorithm is to find the commit rate that reaches the maximum reward, i.e., the minimum time to converge to a certain loss.

\renewcommand{\algorithmicrequire}{\textbf{Parameter Server:}}
\renewcommand{\algorithmicensure}{\textbf{End System:}}
\begin{algorithm}[!t]
\caption{ADSP: Worker and PS Procedures}
\label{algorithm:design}
\begin{algorithmic}[1]
    \Ensure i = 1, 2, ..., m 
    \Function {\textsc{Worker}}{}
        \For {epoch $e = 1, 2, \ldots$}
            \State {receive $\Delta C_{target}^i$ from the scheduler}
            \State set a timer with a timeout of $\frac{\Gamma}{\Delta C_{target}^i} - \mathcal{O}_i$ and invoking TimeOut() upon timeout 
            
			\While {model not converged}
                \State train a minibatch to obtain gradient $g_{i}$
                \State accumulated gradient $U_{i} \leftarrow U_{i} + \eta^{\prime} g_{i}$ ($\eta^{\prime}$ is the local learning rate)
            \EndWhile
        \EndFor
    \EndFunction
    \Function {\textsc{TimeOut}}{}
        \State commit $U_{i}$ to the PS
        \State receive updated global model parameters from the PS and update local model accordingly
        \State restart the timer with timeout of $\frac{\Gamma }{\Delta C_{target}^i} - \mathcal{O}_i$
    \EndFunction
\end{algorithmic}

\begin{algorithmic}[1]
    \Require
    \hspace*{0.05in}
    \Function {\textsc{ParameterServer}}{}
        \While {model not converged}
            \If {receive commit $U_{i}$ from worker $i$}         
                \State $W \leftarrow W - \eta U_{i}$
                \State Send $W$ to worker $i$
            \EndIf
        \EndWhile
    \EndFunction
\end{algorithmic}
\end{algorithm}
\subsection{Worker and PS Procedures} \label{section:c2D_formula}
The procedures at each end system (i.e., worker) and the PS with ADSP is summarized in Alg.~\ref{algorithm:design}, where $\mathcal{O}_i$ represents the communication time for worker $i$ to commit an update to the PS and pull the updated parameters back. At each worker, we use a timer to trigger commit of local accumulative model update to the PS asynchronously, once every $\frac{\Gamma}{\Delta C_{target}^i} - \mathcal{O}_i$ time.

\subsection{Convergence Analysis} \label{section:proof}

We show that ADSP in Alg.~\ref{algorithm:design} ensures model convergence. We define $f_t(W)$ as the objective loss function at step $t$ with global parameter state $W$, where $t$ is the global number of steps (i.e., cumulative number of training steps carried out by all workers). Let $\tilde{W}_t$ be the set of global parameters obtained by ADSP right after step \textit{t}, and $W^{*}$ denote the optimal model parameters that minimize the loss function. We make the following assumptions on the loss function and the learning rate, which are needed for our convergence proof, but are not followed in our experimental settings.

\noindent \textbf{Assumptions:} 
\textit{
\begin{enumerate}[label=(\arabic*), leftmargin=20pt]
  \item $f_t(W)$ is convex
  \item $f_t(W)$ is $L$-Lipschitz, i.e., $\left\|\nabla f_{t}\right\| \leqslant L$
  \item The learning rate decreases as $\eta_t = \frac{\eta}{\sqrt{t}}$,  $t = 1, 2, \ldots$, where $\eta$ is a constant.
\end{enumerate}
}

Based on the assumptions, we have the following theorem on training convergence of ADSP.
\begin{theorem}[Convergence]\label{convergence1}
ADSP ensures that by each checkpoint, the numbers of update commits submitted by any two different workers $i_1$ and $i_2$ are roughly equal, i.e., $c_{i_1} \approx c_{i_2}$. The regret $R = \sum_{t = 1}^{T}f_t(\tilde{W}_t) - f(W^{*})$ is upper-bounded by $O(\sqrt{T})$, when $T \to +\infty$.
\end{theorem}

The {\em regret} is the accumulative difference between the loss achieved by ADSP and the optimal loss over the training course. When the accumulative difference is under a sub-linear bound about \textit{T} (where \textit{T} is the total number of parameter update steps at the PS), we have $f_t(\tilde{W}_t)\to f(W^{*})$ when \textit{t} is large. Then $R/T \to 0$ as $T \to +\infty$, showing that our ADSP model converges to the optimal loss. The detailed proof is given in the supplemental file.

\section{Performance Evaluation}
\label{sec:evaluation}

We implement ADSP as a ready-to-use Python library based on TensorFlow \cite{2016abadi-tensorflow}, and evaluate its performance with testbed experiments.

\subsection{Experiment Setup}
\noindent \textbf{Testbed.} We emulate heterogeneous edge systems following the distribution of hardware configurations of edge devices in a survey \cite{smartphone2018}, using 19 Amazon EC2 instances \cite{amazon}: 7 $\mathtt{t2.large}$ instances, 5 $\mathtt{t2.xlarge}$ instances, 4 $\mathtt{t2.2xlarge}$ instances and 2 $\mathtt{t3.xlarge}$ instances as workers, and 1 $\mathtt{t3.2xlarge}$ instance as the PS.

\noindent \textbf{Applications.} 
We evaluate ADSP with three distributed ML applications: (i) image classification on Cifar-10 \cite{cifar10} using a CNN model from the TensorFlow tutorial \cite{tf_cifar10_tutorial}; (ii) Fatigue life prediction of bogies on high-speed trains, training a recurrent neural network (RNN) model with the dataset collected from the China high-speed rail system; (iii) Coefficient of Performance (COP) prediction of chillers, training a global linear SVM model with a chiller dataset. 

\noindent \textbf{Baselines.} (1) \textit{SSP} \cite{2013ssp}, which allows the fastest worker to run ahead of the slowest worker by up to $s$ steps; (2) \textit{BSP} \cite{1990-bridging}, where the PS strictly synchronizes among all workers such that they always perform the same number of training steps. (3) {\em ADACOMM} \cite{wang2018adaptive}, which allows all workers to accumulate $\tau$ updates before synchronizing with the PS and reduces $\tau$ periodically. (4) {\em Fixed ADACOMM}, a variant of ADACOMM with $\tau$ fixed for all workers. 

\noindent \textbf{Default Settings.}
By default, each mini-batch in our model training includes 128 examples. The check period of ADSP is 60 seconds, and each epoch is 20 minutes long. The global learning rate is $1/M$ (which we find works well through experiments). The local learning rate is initialized to 0.1 and decays exponentially over time.

\begin{figure}[!t]
\begin{center}
	\includegraphics[width = .95\columnwidth]{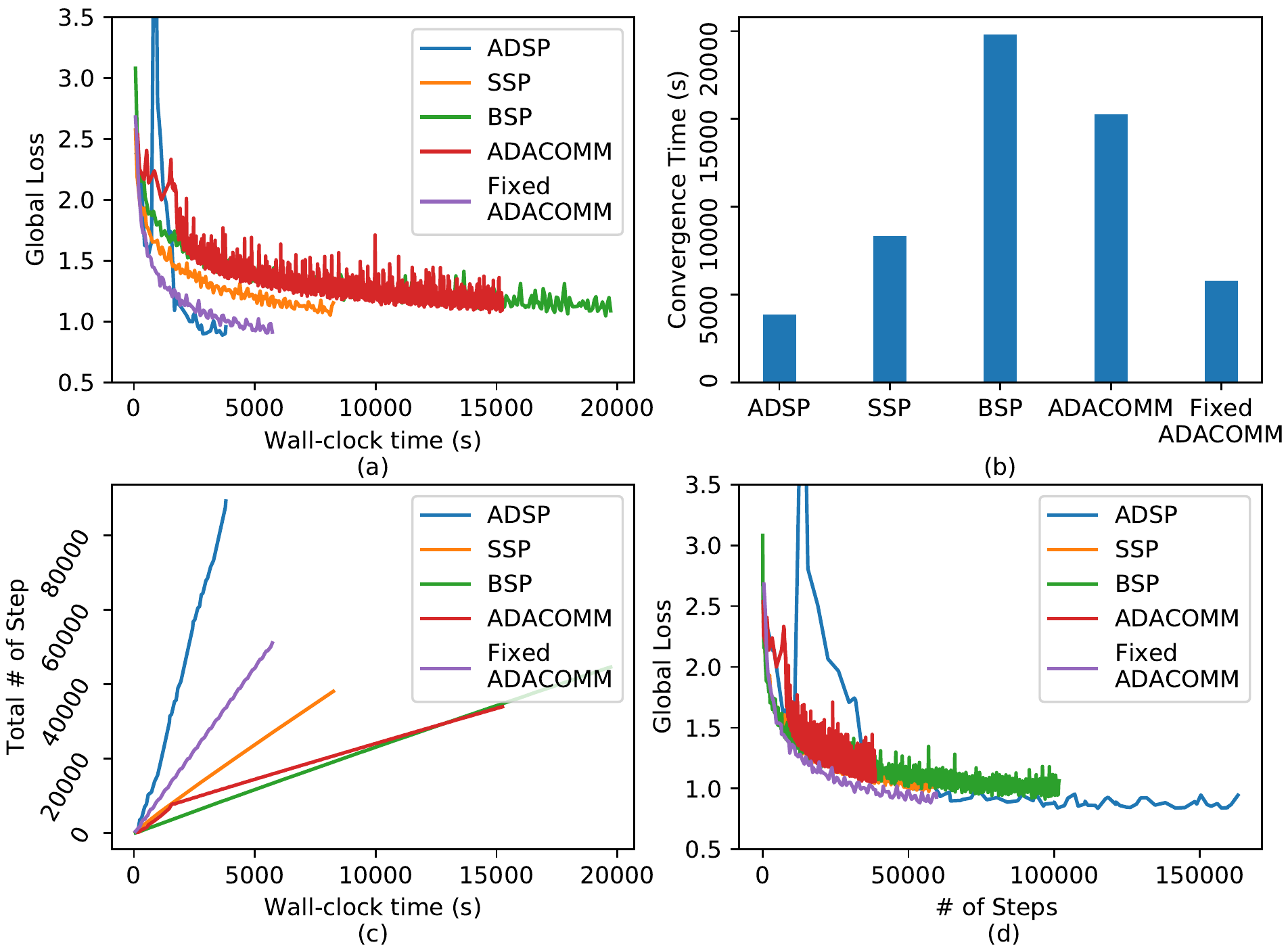}
	\caption{Comparison of ADSP with baselines in training efficiency: training CNN on the Cifar-10 dataset.}
	\label{figure:strain_eval}
\end{center}
\end{figure}

\subsection{Experiment Results}
All results given in the following are based on CNN training on the Cifar-10 dataset. More experiment results on fatigue life prediction  and CoP prediction are given in the supplemental file.

\subsubsection{Performance of ADSP.} \label{section:eval_adsp}
We compare ADSP with the baselines in terms of the wall-clock time and the number of training steps needed to reach model convergence, to validate the effectiveness of no-waiting training of ADSP. In Fig.~\ref{figure:strain_eval}, the global loss is the loss evaluated on the global model on the PS, and the number of steps is the cumulative number of steps trained at all workers. We stop training, i.e., decide that the model has converged, when the loss variance is smaller than a small enough value for 10 steps. Fig.~\ref{figure:strain_eval}(a) plots the loss curves and Fig.~\ref{figure:strain_eval}(b) correspondingly shows the convergence time with each method. We see that ADSP achieves the fastest convergence: $80\%$ acceleration as compared to BSP, $53\%$ to SSP, and $33\%$ to Fixed ADACOMM. For ADACOMM, although we have used the optimal hyper-parameters as in \cite{wang2018adaptive}, it converges quite slowly, which could be due to its instability in tuning $\tau$: $\tau$ is tuned periodically based on the current loss; if the loss does not decrease, it simply multiplies $\tau$ with a constant. In Fig.~\ref{figure:strain_eval}(c), we see that ADSP carries out many more training steps within its short convergence time, which may potentially lead to a concern on its training efficiency.  Fig.~\ref{figure:strain_eval}(d) further reveals that the per-training-step loss decrease achieved by ADSP is slightly lower than that of Fixed ADACOMM, and better than other baselines. The spike in ADSP curve at the beginning stage is due to small commit rates that our search algorithm derives, which make the loss fluctuates significantly. However, with ADSP, the model eventually converges to a smaller loss than losses that other baselines converge to. 

\begin{figure}[!t]
\begin{center}
	\includegraphics[width = .95\columnwidth]{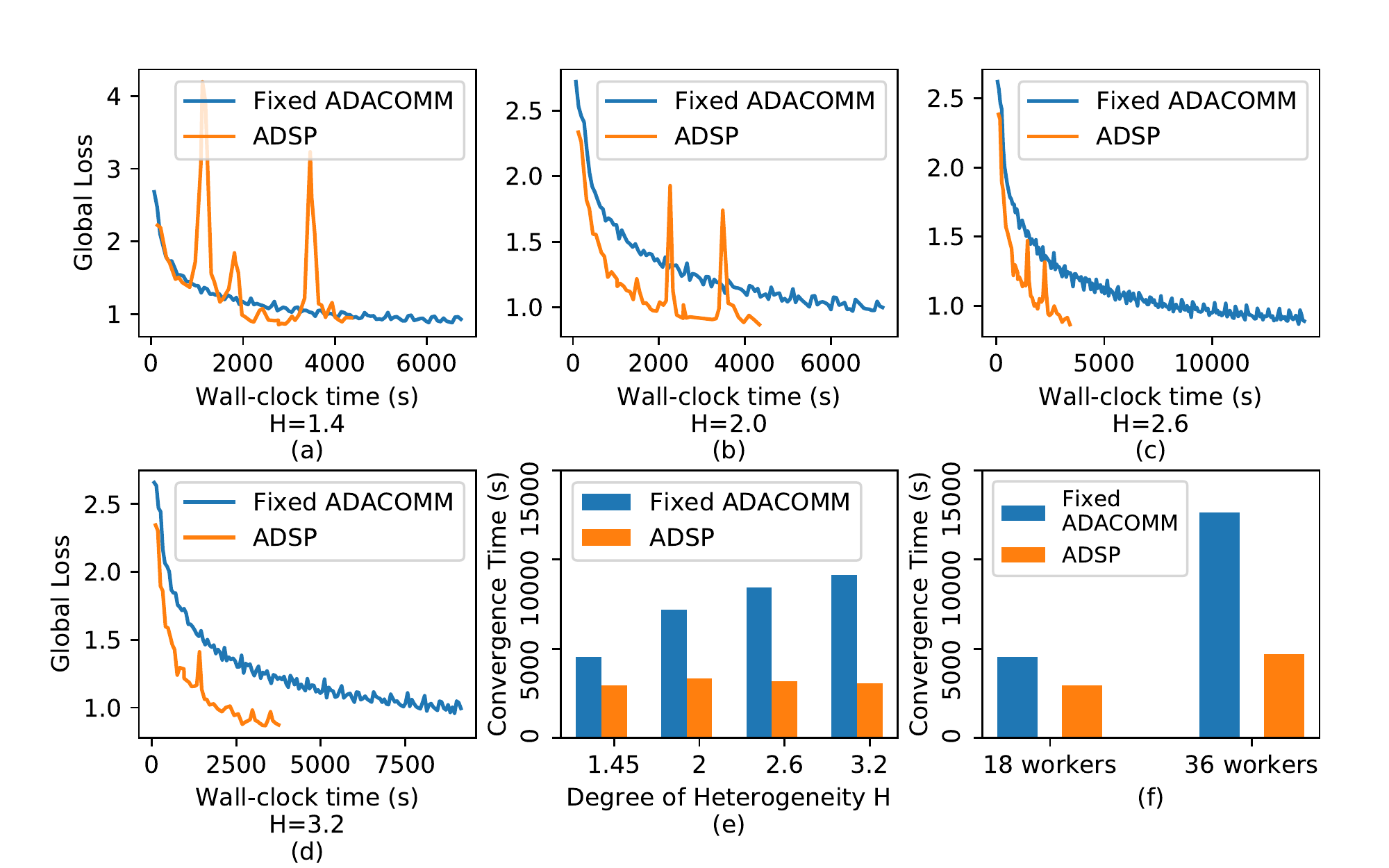}
	\caption{Comparison of ADSP with Fixed ADACOMM at different degrees of heterogeneity and  system scales.}
	\label{figure:vary_heterogeneity}
\end{center}
\end{figure}

\subsubsection{Adaptability to Heterogeneity.}
We next evaluate ADSP's adaptability to different levels of end system heterogeneity. Besides hardware configuration difference among the workers, we further enable each worker to sleep for a specific short time after each step of training one mini-batch, and tune the sleep time to adjust training speeds of workers. We define the heterogeneity degree among the workers as follows:
$$ 
H = \frac{\sum_i^M v_i/M}{\min_{i=1,\ldots,M} v_i} 
$$
where $v_i$ is the number of mini-batches that  worker $i$ can process per unit time. The discussion of the heterogeneity degree considering communication overhead is given in our supplemental file.

Since BSP, SSP and ADACOMM are significantly slower than ADSP in training convergence, here we only compare ADSP with Fixed ADACOMM. Fig.~\ref{figure:vary_heterogeneity}(a)-(d) show that ADSP achieves faster convergence than Fixed ADACOMM (though with more spikes) in different heterogeneity levels. The corresponding convergence times are summarized in Fig.~\ref{figure:vary_heterogeneity}(e), which shows that the gap between ADSP and Fixed ADACOMM becomes larger when the workers differ more in training speeds. ADSP achieves a $62.4\%$ convergence speedup as compared to Fiexd ADACOMM when $H=3.2$. The reason lies in that Fixed ADACOMM still enforces faster workers to stop and wait for the slower workers to finish $\tau$ local updates, so the convergence is significantly influenced by the slowest worker. With ADSP, the heterogeneity degree hardly affects the convergence time much, due to its no-waiting strategy. Therefore, ADSP can adapt well to heterogeneity in end systems.

\subsubsection{System Scalability}\label{sec:scaling}
We further evaluate ADSP with 36 workers used for model training, whose hardware configuration follows the same distribution as in the 18-worker case. Fig.~\ref{figure:vary_heterogeneity}(f) shows that when the worker number is larger, both ADACOMM and ADSP become slower, and ADSP still achieves convergence faster than Fixed ADACOMM (which is more obvious than in the case of smaller worker number). Intuitively, when the scale of the system becomes larger, the chances increase for workers to wait for slower ones to catch up, resulting in that more time being wasted with Fixed ADACOMM; ADSP can use this part of time to do more training, and is hence a more scalable solution in big ML training jobs.

 \subsubsection{The Impact of Network Latency.}\label{section:latency}

\begin{figure}
\begin{center}
  \includegraphics[width = .95\columnwidth]{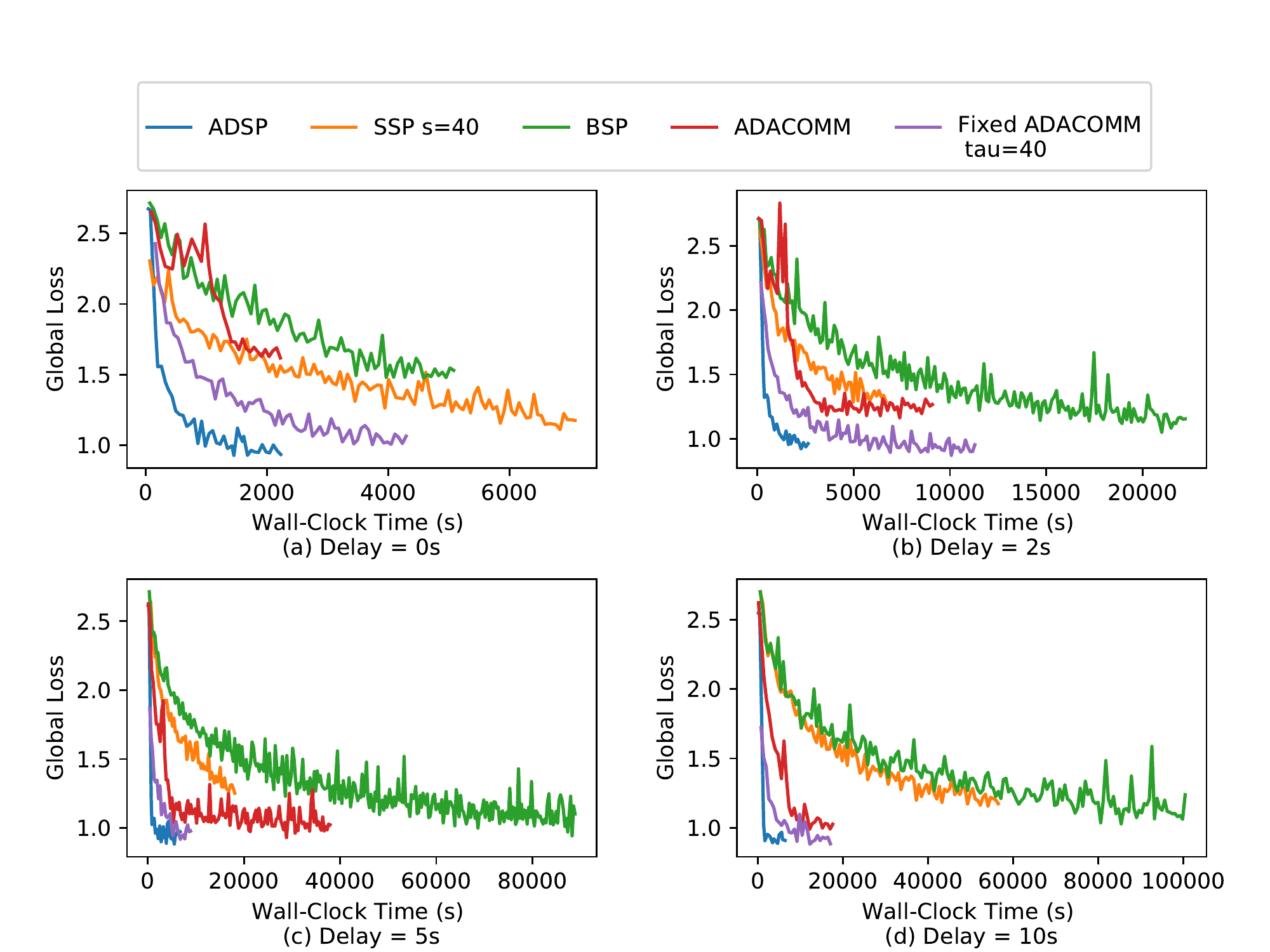}
  \caption{Comparison of ADSP with baselines with different network delays.}
  \label{figure:delay}
\end{center}
\end{figure}

Edge systems usually have relatively poor network connectivity \cite{konevcny2016federated}; the communication time for each commit is not negligible, and could be even larger than the processing time in each step. Fig.~\ref{figure:delay} presents the convergence curve of each method as we add different extra delays to the communication module. When we increase the communication delay, the speed-up ratio of {\em ADSP}, {\em Adacomm} and {\em Fixed Adacomm}, as compared to {\em BSP} and {\em SSP}, becomes larger. This is because the first three models allow local updates and commit to the PS less frequently, consequently less affected by the communication delay than the last two methods. Among the first three models, ADSP still performs the best in terms of convergence speed, regardless of the communication delay. 
  
 The rationale behind is that we can count the communication time when evaluating a worker's `\textit{processing capacity}': for worker $i$, the average processing time per training step is $t_i + \mathcal{O}_i/\tau_i$, where $t_i$ is the time to train a mini-batch, $\mathcal{O}_i$ is the communication time for each commit, and $\tau_i$ is the number of local updates between two commits.Therefore, we can extend the scope of heterogeneity in processing capacity to include the heterogeneity of communication time as well. ADSP only needs to ensure the commit rates of all workers are consistent, and can inherently handle the generalized heterogeneity without regard to which components cause the heterogeneity.
 
\section{Concluding Remarks}
This paper presents ADSP, a new parameter synchronization model for distributed ML with heterogeneous edge systems. ADSP allows workers to keep training with minimum waiting and enforces approximately equal numbers of commits from all workers to ensure training convergence. An online search algorithm is carefully devised to identify the near-optimal global commit rate. ADSP maximally exploits computation resources at heterogeneous workers, targeting training convergence in the most expedited fashion. Our testbed experiments show that ADSP achieves up to $62.4\%$  convergence acceleration as compared to most of the state-of-the-art parameter synchronization models. ADSP is also well adapted to different degrees of heterogeneity and large-scale ML applications.

\section{Acknowledgements}
This work was supported in part by grants from Hong Kong RGC under the contracts HKU 17204715, 17225516, C7036-15G (CRF), C5026-18G (CRF), in part by WHU-Xiaomi AI Lab, and in part by GRF PolyU 15210119, ITF UIM/363, CRF C5026-18G, PolyU 1-ZVPZ, and a Huawei Collaborative Grant.

\bibliographystyle{aaai}
\bibliography{biblio}

\clearpage
\begin{appendices}
\section{Proof of Theorem 1 }\label{appendix:local_update}
We use $t$ to index the state of the global parameter state. $W_t$ denotes the state of parameters after the $t^{th}$ commit, $\eta$ is the global learning rate, Let $U(W_t)$ denote the accumulative local updates (i.e., gradients) that a worker commits when the global model is $W_t$. To prove Theorem 1, we explain that how the gloabl parameter update equation of ADSP is equivalent to the following form.

$$\mathbb{E}(W_{t+1} - W_{t}) = (1-p)\mathbb{E}(W_{t} - W_{t-1}) - p \eta \mathbb{E} U(W_t)$$

We assume the time that it cost to accumulate local updates for commit $t$ obeys the exponential distribution $W_t \sim Exp(\lambda)$, where $\lambda$ is the frequency of commits. The staleness distribution $\tau_t$, we define the number of commits from other workers between a worker's two commits as the staleness, which follows a geometrical distribution $\tau_t \sim Geom(p)$. Because we assume each commit is independent, the number of commits from other workers in $T$ units of time is
\vspace{-3mm}
$$B_t(T) \sim \mathit{Poisson}(\lambda T (1 - \frac{1}{M}) \sum_{i=1}^{M}\frac{\Gamma}{\Delta C_{target}^i v_i})$$
So the staleness should be euqal to the number of commits from other workers in $W_t$ units of time, i.e., $\tau_t \sim B_t(W_t)$.
According to \cite{Poisson}, We can convert this combination to a geometrical distribution

\vspace{-3mm}
\begin{equation}
	\begin{aligned}
	& \tau_t \sim Geom(p), \quad p = \frac{M}{M + (M - 1)\sum_{i=1}^{M}\frac{\Gamma}{\Delta C_{target}^i v_i}} \\
	& 1 - p = 1 - \frac{1}{1 + (1 - \frac{1}{M}) \sum_{i=1}^{M}\frac{\Gamma}{\Delta C_{target}^i v_i}}
	\end{aligned}
\end{equation}
\vspace{-3mm}

Suppose the noisy global parameters update euqation can be written as
\begin{equation}
\begin{aligned}
W_{t+1} = W_t - \eta U(\tilde{W}^t)
\end{aligned}
\end{equation}
where $\tilde{W}^t$ is the real state of parameters on which the $t$'th commit is calculated and is uncertain. Considering the staleness of each commit follows a geometrical distribution, let $Prob(\tau_t = l) = p(1-p)^l = P_l$,  we can get the expectation of $W_{t+1}$.
\begin{equation}
\begin{aligned}
\mathbb{E}[W_{t+1}] = \mathbb{E}[W_t] - \eta \left(\sum_{l=0}^{\infty}P_l \mathbb{E}U(W_{t-l})\right)
\end{aligned}
\end{equation}
Then, we can get the expectation of $W_{t+1}-W_{t}$

\small{
\begin{equation}\label{eq:proof_result}
\begin{aligned}
\quad & \mathbb{E}[W_{t+1} - W_{t}] \\
= & \mathbb{E}[W_t - W_{t-1}] - \\
	& \eta\left(\sum_{l=0}^{\infty}P_l \mathbb{E}U(W_{t-l}) - \sum_{l=0}^{\infty}P_l \mathbb{E}U(W_{t-1-l})\right) \\
= & \mathbb{E}[W_t - W_{t-1}] - \eta P_0 \mathbb{E}U(W_{t}) - \\
	& \eta \sum_{l=0}^{\infty} (P_{l+1} - P_l)\mathbb{E}U(W_{t-1-l}) \\
= & \mathbb{E}[W_t - W_{t-1}] - \eta p \mathbb{E}U(W_{t}) - \eta \sum_{l=0}^{\infty} (p(1-p)^{l+1} \\
& - p(1-p)^l)\mathbb{E}U(W_{t-1-l}) \\
= & \mathbb{E}[W_t - W_{t-1}] - \eta p \mathbb{E}U(W_{t}) + \\
	& \eta p\sum_{l=0}^{\infty} p(1-p)^l\mathbb{E}U(W_{t-1-l}) \\
= & \mathbb{E}[W_t - W_{t-1}] - \eta p \mathbb{E}U(W_{t}) - p\sum_{l=0}^{\infty} \mathbb{E}[W_t - W_{t-1}] \\
= & (1-p)\mathbb{E}[W_t - W_{t-1}] - \eta p \mathbb{E}U(W_{t})
\end{aligned}
\end{equation}
}
Eqn. (\ref{eq:proof_result}) can be regarded as the momentum-based SGD update equation (\ref{eq:sgdMomentum2}), thus we prove that the asynchrony induced by local updates can be converted to an implicit momentum term.
\begin{equation}\label{eq:sgdMomentum2}
\begin{aligned}
W_{t+1} - W_t = \mu (W_{t} - W_{t-1}) - \eta \nabla \ell(W_t)
\end{aligned}
\end{equation}

\section{Convergence Proof Under ADSP}\label{appendix:ADSP}
We follow the similar mehthod as that used in SSP\cite{2013ssp} to formularize our ADSP model. In our proof, we focus on \textit{Stochastic Gradient Descent(SGD)}, 
which is very famous and widly used for finding the optimal parameters in the machine learning model. Our final analysis result is as below,

Suppose there are \textit{m} workers which push updates to the parameter server at different time intervals. Take worker \textit{i} for instance, $\tilde{W}_{c, i, k} $ denotes the noisy local state of parameters in worker \textit{i} after \textit{c commits} and \textit{k local updates}, and let $\textbf{u}_{c, i, k}$ denote the $k_{th}$ local undate of worker \textit{i} after \textit{c} commits. And $W_0$ is the initial state of parameters in each worker. When we get $\tilde{W}_{c, i, k} $ defined above at some point, let $c_{c, i, k}^{i^{'}}$ represent the number of \textit{commits} of worker $i^{'}$ at this point. For simplicity, we use $c^{i^{'}}$ instead of $c_{c, i, k}^{i^{'}}$ in the following analysis. 

So the noisy state is equal to:
$$\tilde{W}_{c, i, k} = W_0 + \sum_{k^{'} = 1}^{k}\textbf{u}_{c, i, k^{'}} + \sum_{i^{'} = 1}^{m}\sum_{c^{'} = 0}^{c^{i^{'}}}\sum_{k^{'} = 1}^{\Delta_{i^{'}}}\textbf{u}_{c^{'}, i^{'}, k^{'}}$$

We define a reference sequence for state $W_t$ as follows, to link the noisy state sequence and optimal state sequence together. The reference sequence is viewed as the `true' sequence of parameters produced by forcing workers to commit in a Round-Robin order. In other words, we can consider the reference sequence of state $W_t$ as that generated by training in the order as described in Algorithm \ref{algorithm:reference}.

$$ W_t = W_0 + \sum_{t^{'} = 0}^{t}\textbf{u}_t' = W_0 + \sum_{t^{'} = 0}^{t}\textbf{u}_{c, i, k} $$
where for each $t^{'} $, 
$$ s.t., \quad t^{'} = c \cdot \sum_{i^{'} = 1}^{m}D_{i^{'}} + \sum_{i^{'} = 1}^{i - 1}D_{i^{'}} + k, $$

Here $D_i$ denotes the number of local updates between two commits of worker $i$, and its value can be calculated based on the commit rate $\Delta C_{target}^i$ and the speed of this machine $v_i$, by taking the following transform euqation.
$$D_i = \frac{\Gamma}{\Delta C_{target}^i * v_i}$$
where $\Gamma$ denotes the check period.

\renewcommand{\algorithmicrequire}{\textbf{Initialization:}}
\renewcommand{\algorithmicensure}{\textbf{Defination:}}
\begin{algorithm}  
  \caption{Reference sequence of state} 
  \label{algorithm:reference}
  \begin{algorithmic}[1]  
  \Ensure \textit{m} is the total number of workers,
        $D_{i}$ is the commit interval for worker \textit{i}.
    \For {c = 1, 2, ...}
       \For {i = 1, 2, ...., m}
         	\For {k = 1, 2, ..., $D_{i}$}
      			\State train an example or a mini-batch examples
      			\State calculate an update
        	\EndFor
        	\State commit accumulative updates
      \EndFor
    \EndFor
  \end{algorithmic}  
\end{algorithm} 

Then we can derive that the reference state $W_t$(i.e., $W_{c, i, k}$) is equal to:
\begin{equation}
\begin{aligned}
W_{c, i, k} = W_0 + \sum_{k^{'} = 1}^{k}\textbf{u}_{c, i, k^{'}} + \sum_{i^{'} = 1}^{m}\sum_{c^{'} = 1}^{c}\sum_{k^{'} = 1}^{D_{i^{'}}}\textbf{u}_{c^{'}, i^{'}, k^{'}}
\end{aligned}		      
\end{equation}
	
We can get the relationship between the noisy sequence of state and the reference one as below:
$$\tilde{W}_{t} = W_{t}  - \sum_{i \in \mathcal{A}_t}\textbf{u}_i + \sum_{i \in \mathcal{B}_t}\textbf{u}_i $$
where we have decomposed the difference between the nosiy state $\tilde{W}_{t}$ and the reference state $W_{t}$ into $\mathcal{A}_t$, the index set of updates $\textbf{u}_i$ that are missing from $W_{t}$, and $\mathcal{B}_t$, the index set of ``extra'' updates in $\tilde{W}_{t}$ but not in $W_{t}$. 

Again, we reclaim Theorem 1 here.
\begin{theorem}[Convergence]\label{convergence2}
Suppose ADSP is used when training an ML model with SGD, where the objective loss function $f$ is convex and the learning rate decreases as $\eta_t = \frac{\eta}{\sqrt{t}}$,  $t = 1, 2, \ldots$, where $\eta$ is a constant. Let $\tilde{W}_t$ be the set of global parameters obtained by ADSP right after step \textit{t}, and $W^{*}$ denote the ideal optimal parameters that can minimize the loss function. ADSP ensures that between any two different workers $i_1$ and $i_2$, the commit numbers by each checkpoint is roughly equal, i.e., $\|c_{i_1} - c_{i_2}\| \leq \epsilon $, where $\epsilon$ is a small constant, and the regret $R = \sum_{t = 1}^{T}f_t(\tilde{W}_t) - f(W^{*})$ is bound by $O(\sqrt{T})$, when $T \to +\infty$.
\end{theorem}

\textbf{Proof: }
We define $D(W \| W^{\prime}) := \frac{1}{2} \| W - W^{\prime} \|^2$, and assume that for any $W, W^{\prime} \in X, D(W \| W^{\prime}) \leqslant F^2$, where $X$ is the possible value space of parameters, i.e.,the optimization problem has bounded diameter. Because we use SGD as our algorithm, our goal is to optimize the \textit{convex} objective function \textit{f} by iteratively applying gradient descent $\bigtriangledown f_t$. So $f_t(W_1) - f_t(W_2) \leqslant \left \langle \bigtriangledown f_t(W_1),  W_1 - W_2 \right \rangle$. And we assume that $f_t$ is \textit{L}-Lipschitz, so $\| \bigtriangledown f_t \| \leqslant L$. Let the learning rate $\eta_t = \frac{\eta}{\sqrt{t}}$, with t = 1, 2, ..., T. Here $F, \eta$ and $L$ are all constants.

So, we first get the regret as follows,
\begin{equation}
\begin{aligned}
\label{equation:2}
R   & := \sum_{t = 1}^{T}f_t(\tilde{W}_t) - f_t(W^{*}) \\
    & \leqslant \sum_{t = 1}^{T} \left \langle \bigtriangledown f_t(\tilde{W}_t), \tilde{W}_t - W^{*} \right \rangle \\
    & = \sum_{t = 1}^{T} \left \langle \tilde{\textbf{g}}_t, \tilde{W}_t - W^{*} \right \rangle
\end{aligned}         
\end{equation}
where $\tilde{\textbf{g}}_t$ denotes $\bigtriangledown f_t(\tilde{W}_t)$.

\textbf{Lemma 1: }
For any $\tilde{W}_t, W^{*} \in \mathbb{R}^n$,
\begin{equation}
\begin{aligned}
\label{equation:3}
 & \left \langle \tilde{\textbf{g}}_t, \tilde{W}_t - W^{*} \right \rangle =  \frac{1}{2} \eta_t\| \tilde{\textbf{g}}_t \|^2 \\
 & + \frac{D(W^{*} \| W_{t}) - D(W^{*} \| W_{t + 1})}{\eta_t} \\
 & + \sum_{t^{\prime} \in \mathcal{A}_t} \eta_{t^{\prime}}\left \langle \tilde{\textbf{g}}_{t^{\prime}}, \tilde{\textbf{g}}_t \right \rangle 
	- \sum_{t^{\prime} \in \mathcal{B}_t} \eta_{t^{\prime}}\left \langle \tilde{\textbf{g}}_{t^{\prime}}, \tilde{\textbf{g}}_t \right \rangle
\end{aligned}		      
\end{equation}

\textbf{Proof: }
\begin{equation}
\begin{aligned}
\label{equation:4}
& D(W^{*} \| W_{t + 1}) - D(W^{*} \| W_{t}) =  \frac{1}{2} \| W^{*} - W_{t} + W_{t} - W_{t + 1} \| ^2 \\
		& \quad - \frac{1}{2} \| W^{*} - W_{t} \| ^2 \\
&	= \frac{1}{2} \| W^{*} - W_{t} + \eta_{t} \tilde{\textbf{g}}_{t} \|^2 - \frac{1}{2} \| W^{*} - W_{t} \| ^2  \\
&	= \frac{1}{2}\eta_t^2 \| \tilde{\textbf{g}}_t \|^2 - \eta_{t} \left \langle W_{t} - W^{*}, \tilde{\textbf{g}}_{t}\right \rangle \\
&	= \frac{1}{2}\eta_t^2 \| \tilde{\textbf{g}}_t \|^2 - \eta_{t} \left \langle \tilde{W}_t - W^{*} +  W_{t} - \tilde{W}_t , \tilde{\textbf{g}}_{t}, \right \rangle \\
&	= \frac{1}{2}\eta_t^2 \| \tilde{\textbf{g}}_t \|^2 - \eta_{t} \left \langle \tilde{W}_t - W^{*}, \tilde{\textbf{g}}_{t}\right \rangle\\
& \quad + \eta_{t} \left \langle W_{t} - \tilde{W}_t, \tilde{\textbf{g}}_{t}\right \rangle 
\end{aligned}		      
\end{equation}

Where $\beta_j^t$ denotes a coefficient according to the class of example at \textit{t} step. And 
\begin{equation}
\begin{aligned}
\label{equation:5}
& \left \langle W_{t} - \tilde{W}_t, \tilde{\textbf{g}}_{t}\right \rangle \\
& = \left \langle \left[ - \sum_{t^{\prime} \in \mathcal{A}_t} \eta_{t^{\prime}} \tilde{\textbf{g}}_{t^{\prime}}  + \sum_{t^{\prime} \in \mathcal{B}_t} \eta_{t^{\prime}}  \tilde{\textbf{g}}_{t^{\prime}} \right], \tilde{\textbf{g}}_t \right \rangle \\
& =  - \sum_{t^{\prime} \in \mathcal{A}_t} \eta_{t^{\prime}}\left \langle \tilde{\textbf{g}}_{t^{\prime}}, \tilde{\textbf{g}}_t \right \rangle 
	+  \sum_{t^{\prime} \in \mathcal{B}_t} \eta_{t^{\prime}}\left \langle \tilde{\textbf{g}}_{t^{\prime}}, \tilde{\textbf{g}}_t \right \rangle
\end{aligned}		      
\end{equation}
Combining Equation \ref{equation:4} and Equation \ref{equation:5}, complete the proof of Lemma 1
 
 \textbf{Back to Equation \ref{equation:2}: } We use Lemma 1 to continue the proof:
\begin{equation}
\begin{aligned}
\label{equation:6}
R & \leqslant \sum_{t = 1}^{T} \left \langle \tilde{\textbf{g}}_t, \tilde{W}_t - W^{*} \right \rangle \\
	= &  \sum_{t = 1}^{T} \frac{1}{2} \eta_t \| \tilde{\textbf{g}}_t \|^2 +  \sum_{t = 1}^{T}\frac{D(W^{*} \| W_{t}) - D(W^{*} \| W_{t + 1})}{\eta_t} \\
	& +  \sum_{t = 1}^{T} \left[ \sum_{t^{\prime} \in \mathcal{A}_t} \eta_{t^{\prime}}\left \langle \tilde{\textbf{g}}_{t^{\prime}}, \tilde{\textbf{g}}_t \right \rangle
	-  \sum_{t^{\prime} \in \mathcal{B}_t} \eta_{t^{\prime}}\left \langle \tilde{\textbf{g}}_{t^{\prime}}, \tilde{\textbf{g}}_t \right \rangle \right] \\
	= &  \sum_{t = 1}^{T} \left[ \frac{1}{2} \eta_t \| \tilde{\textbf{g}}_t \|^2 + 
	\sum_{t^{\prime} \in \mathcal{A}_t} \eta_{t^{\prime}}\left \langle \tilde{\textbf{g}}_{t^{\prime}}, \tilde{\textbf{g}}_t \right \rangle \right. \\
	  & \left. - \sum_{t^{\prime} \in \mathcal{B}_t} \eta_{t^{\prime}}\left \langle \tilde{\textbf{g}}_{t^{\prime}}, \tilde{\textbf{g}}_t \right \rangle \right] + \frac{D(W^{*} \| W_{1})}{\eta_1} 
	 - \frac{D(W^{*} \| W_{T+1})}{\eta_T } \\
	& + \sum_{t = 2}^{T} \left[ D(W^{*} \| W_{t})(\frac{1}{\eta_t} - \frac{1}{\eta_{ t-1}}) \right]      
\end{aligned}		      
\end{equation}

First, we first bound the upper limit of term: $\sum_{t = 1}^{T} \frac{1}{2} \eta_t \| \tilde{\textbf{g}}_t \|^2$:
\begin{equation}
\begin{aligned}
\label{equation:7}
\sum_{t = 1}^{T} \frac{1}{2} \eta_t \| \tilde{\textbf{g}}_t \|^2 & \leqslant \sum_{t = 1}^{T} \frac{1}{2} \eta_t L^2 \\
	= & \sum_{t = 1}^{T} \frac{1}{2} \frac{\eta}{\sqrt{t}} L^2 \\
	\leqslant & K \eta L^2 \sqrt{T}
\end{aligned}
\end{equation}

Next, we bound the upper limit of the second term:

\begin{equation}
\begin{aligned}
\label{equation:8}
 & \frac{D(W^{*} \| W_{1})}{\eta_1} - \frac{D(W^{*} \| W_{T+1})}{\eta_T } + \sum_{t = 2}^{T} \left[ D(W^{*} \| W_{t})(\frac{1}{\eta_t} - \frac{1}{\eta_{t-1}}) \right] \\
 & \leqslant \frac{F^2}{\eta} + 0 + \frac{F^2}{\eta} \sum_{t = 2}^{T}\left[\sqrt{t} - \sqrt{t-1} \right] \\
 & = \frac{F^2}{\eta} + \frac{F^2}{\eta} (\sqrt{T} - 1) \\
 & = \frac{F^2}{\eta} \sqrt{T}
\end{aligned}
\end{equation}

Third, before we bound the upper limit of the rest term, we give a lemma:
\\\\
\textbf{Lemma 2:}
For any different worker machine $i_1$ and $i_2$, 
as long as $|c_{i_1} - c_{i_2}| \leqslant \epsilon$ where $\epsilon$ is a constant, 
$|\mathcal{A}_t| + |\mathcal{B}_t| \sim O(1)$ and 
the minimum $t \in (\mathcal{A}_t \bigcup \mathcal{B}_t), t_{min} \sim O(t)$
\\\\
\textbf{Proof:}
\begin{equation}
\begin{aligned}
\label{equation:9}
|\mathcal{A}_t| + |\mathcal{B}_t| = & \sum_{i^{'} = 1}^{i-1} | D_{i^{'}} \cdot (c + 1) - D_{i^{'}} \cdot c^{i^{'}}| + \\
		      & \sum_{i^{'} = i+1}^{P} | D_{i^{'}} \cdot c - D_{i^{'}} \cdot c^{i^{'}}| \\
		   = & \sum_{i^{'} = 1}^{i-1} D_{i^{'}} \cdot | c - c^{i^{'}} + 1| + \\
		      & \sum_{i^{'} = i+1}^{m} D_{i^{'}} \cdot |c - c^{i^{'}}| \\
        \leqslant & \delta \cdot (\sum_{i^{'} = 1}^{m} | c - c^{i^{'}}| + m) \\
        \leqslant & (\delta \epsilon + 1) \cdot m \sim O(1)
\end{aligned}		      
\end{equation}
And because $|c_{i_1} - c_{i_2}| \leqslant \epsilon$ for any worker machine $i_1$ and $i_2$, 
suppose at \textit{t} step, for the minimum $t \in (\mathcal{A}_t \bigcup \mathcal{B}_t), t_{min}$, we have:
$$ t_{min} \geqslant t - \epsilon\sum_{i^{\prime} = 1}^{m}D_{i^{\prime}} \sim O(t)$$

So, we can bound the upper limit of the final term:

\begin{equation}
\begin{aligned}
\label{equation:9}
& \sum_{t = 1}^{T} \left[ \sum_{t^{\prime} \in \mathcal{A}_t} \eta_{t^{\prime}}\left \langle \tilde{\textbf{g}}_{t^{\prime}}, \tilde{\textbf{g}}_t \right \rangle
	-  \sum_{t^{\prime} \in \mathcal{B}_t} \eta_{t^{\prime}}\left \langle \tilde{\textbf{g}}_{t^{\prime}}, \tilde{\textbf{g}}_t \right \rangle \right] \\
\leqslant &  \sum_{t = 1}^{T} \left[ | \mathcal{A}_t | + | \mathcal{B}_t | \right] \eta_{max} L^2 \\
\leqslant & (\delta \epsilon + 1) m \eta L^2 \sum_{t = 1}^{T} \frac{1}{\sqrt{t - \epsilon\sum_{i^{\prime} = 1}^{m}D_{i^{\prime}}}} \\
\leqslant & 2(\delta \epsilon + 1) m \eta L^2 \sqrt{T - \epsilon\sum_{i^{\prime} = 1}^{m}D_{i^{\prime}}} \\
\leqslant & 2(\delta \epsilon + 1) m \eta L^2 \sqrt{T}
\end{aligned}
\end{equation}
Note that $\sum_{i = 1}^{b}\frac{1}{2\sqrt{i}} \leqslant \sqrt{b - a + 1}$.
\\\\
\textbf{Back to Equation \ref{equation:2}: }
We final get the result:
\begin{equation}
\begin{aligned}
\label{equation:10}
& R = \sum_{t = 1}^{T}f_t(\tilde{W}_t) - f_t(W^{*}) \\
	&\leqslant \eta L^2 \sqrt{T} + \frac{F^2}{\eta} \sqrt{T} + 2(\delta \epsilon + 1) m\eta L^2 \sqrt{T}
\end{aligned}
\end{equation}
This completes the proof of Theorem 1.

\begin{table}[htb]
  \centering
  \caption{AMAZON EC2}
  \label{table:aws}
    \begin{tabular}{|c|c|c|c|}%
      \toprule[1.5pt]
       	\textbf{Type} & \textbf{vCPUs} & \begin{tabular}{@{}c@{}}\textbf{Memory} \\ \textbf{(GiB)} \end{tabular} & \begin{tabular}{@{}c@{}}\textbf{Number of} \\ \textbf{Instances}\end{tabular} \\
      	\midrule[1pt]
      	$\mathtt{t2.large}$	& 2 & 8 & 7 \\
	    \hline
      	$\mathtt{t2.xlarge}$	 & 4	& 16 & 5 \\
      	\hline
      	$\mathtt{t2.2xlarge}$ & 8 & 32 & 4 \\
      	\hline
      	$\mathtt{t3.2xlarge}$ & 8 & 32 & 1 \\
      	\hline
      	$\mathtt{t3.xlarge}$	 & 4	& 16 & 2 \\
      \bottomrule[1.5pt]
    \end{tabular}
\end{table}

\begin{table}[htb]
\centering
\caption{Smart Phone Market Share in the USA during over Q2  2018}
\label{table:smartPhone}
\begin{tabular}{|c|c|c|}%
\toprule[1.5pt]
 &	Geekbench   &	Share\\
Vendor &	4.1/4.2 64 Bit  & (USA)	\\
 & Multi-Core Score & \\
\midrule[1pt]
iPhone 6 &	2759 & $6.22\%$ \\
\hline
iPhone 6S &	4459 & $7.77\%$ \\
\hline
iPhone 6S Plus	& 4459 &	$4.34\%$ \\
\hline
iPhone SE &	4459 &	$3.89\%$ \\
\hline
iPhone 7 &	5937	 & $12.05\%$ \\
\hline
iPhone 7 Plus &	5937	 & $9.96\%$ \\
\hline
Samsung Galaxy S8 &	6711 & $2.96\%$ \\
\hline
iPhone 8 Plus & 11421 & $5.68\%$ \\
\hline
iPhone X & 11421	 & $5.00\%$ \\
\hline
iPhone 8 & 11421	 & $4.04\%$ \\
\bottomrule[1.5pt]
\end{tabular}
\end{table}

\section{Convergence Speed Analysis}\label{appendix:speed}
Given $m$ workers, worker $i$ trains $v_i$ steps per second, thus each step needs $t_i = \frac{1}{v_i}$ seconds, $\mathcal{O}_i$ is the communication overhead when worker $i$ sends updates to the PS.

\subsection*{BSP}
The average speed: 
$$ V_{BSP} = \frac{1}{\max_i (t_i + \mathcal{O}_i )} \ (step/s)$$

\subsection*{Fixed Adacomm}
Worker $i$ sends updates to the PS every $\tau$ steps. The average speed: 
$$ V_{Fixed} = \frac{1}{\max_i \tau(t_i + \frac{\mathcal{O}_i}{\tau})} \ (step/s)$$

\subsection*{SSP}
The threshold is $s$, suppose $s = \tau$, the average speed should be:
$$ V_{BSP} \leq V_{SSP} \leq V_{Fixed}$$ 
If $s=1$, $V_{BSP} = V_{SSP}$. If the cluster is homogeneous, $V_{SSP} = V_{Fixed}$.

\subsection*{ADSP}
Define $T_{checkperiod}$ as the check period, and the number of commits during one check period is $\Delta C_{target}$. The average speed is:
$$ V_{ADSP} = \frac{1}{m}\sum_{i}^{m} \frac{1}{t_i + \mathcal{O}_i / \tau_i} \ (step/s)$$ 
$$ s.t. \quad t_i \tau_i + \mathcal{O}_i = \frac{T_{checkperiod}}{\Delta C_{target}}, i = 1, 2, ..., m$$

\subsection*{Conclusion}
We observe that the average speed of each system is a function of $t_i + \mathcal{O}_i/\tau_i$ (for BSP, $\tau_i$ = 1), thus we can regard $t_i^\prime = t_i + \mathcal{O}_i/\tau_i$ as the time worker $i$ needs to train a step and assume the communication overhead is 0. In this way, we extend the concept of heterogeneity to include the heterogeneity of communication time.

\ignore{
\begin{table*}[htb]
  \centering
  \caption{Distributed ML frameworks}
  \label{table_notation}
    \begin{tabular*}{\linewidth}{|c|c|c|c|}%
      \toprule[1.5pt]
       	\textbf{Platform}	& \textbf{BSP} 				& \textbf{SSP} 				& \textbf{ADSP }		\\
      	\midrule[1pt]
      	Heterogeneous		&  \XSolidBrush		 & \makecell[tl]{Can adapt to \\slight heterogeneity}& \makecell[tl]{Can adapt to \\sever heterogeneity}	\\
	\hline
      	Data Imbalance		& \XSolidBrush	 	& \XSolidBrush				& \Checkmark			\\
	\hline
        	Bandwidth			& \Hu{?}			& \Hu{?}					& \Hu{?}	\\
	\hline
	Model Size		& \makecell[tl]{Adapt to large models} & \makecell[tl]{Adapt to large models } 	& \makecell[tl]{Perform better \\than SSP and BSP \\when model is \\small}	\\
	\hline
	Scalability			& \makecell[tl]{Adapt to small-scale \\clusters} & \makecell[tl]{Adapt to small-scale \\clusters}		& \makecell[tl]{Adapt to large-scale \\clusters}	\\
	\hline
	\multicolumn{4}{|c|}{\makecell[tl]{
	\Hu{Scalability} is related to heterogeneity, \\
	if the scale become larger and there is heterogeneity, ADSP tends to perform better than SSP. \\
	But if the cluster is still homogeneous, which framework is better is not sure.\\ 
	\\
	\Hu{Bandwidth.}According to theoretical analysis, suppose in \STrain system, \\
	each worker commits to the PS every $\theta$ seconds, \\
	the slowest machine's training speed is $v_{min}$, if $\theta > \frac{s}{v_{min}}$, ADSP is faster.}} \\
      \bottomrule[1.5pt]
    \end{tabular*}
\end{table*}

\clearpage
\begin{table*}[htb]
  \centering
  \begin{minipage}[t]{0.9\linewidth} 
  \caption{Notation Table}
  \label{table_notation}
    \begin{tabular*}{\linewidth}{ll}%
      \toprule[1.5pt]
       	\textbf{Notation} & \textbf{Description} \\
      \midrule[1pt]
      	m & total number of workers \\
      	i & the index of workers \\
        j & the index of class \\
        t & the index of step \\
      	$D_i$	& the accumulative number of updates between two pushes for worker \textit{i}\\
	$\delta$ 	& the upper limit of $D_i$ \\
      	{$\tilde{W}_{c, i, k} $} 	&  the noisy local state of parameters in worker \textit{i} after \textit{c commits} and \textit{k local updates} \\
      	$\textbf{u}_{c, i, k}$ 		& the $\textit{k}_{th}$ local update in worker \textit{i} after \textit{c commits} \\
      	$W_0$ 			& the initial state of parameters in each worker \\
      	$c_{c, i, k}^{i^{'}}, $(i.e. $c^{i^{'}}$ ) 		& the number of \textit{commits} of worker $i^{'}$ at the point when we get $\tilde{W}_{c, i, k} $  \\
      	$\mathcal{A}_t$ 	& the index set of updates  $\textbf{u}_i$ that are missing from $W_{t}$ \\
        $\mathcal{B}_t$	  & the index set of "extra" updates in $\tilde{W}_{t}$ but not in $W_{t}$ \\
        $c_{i}$ 	       & total number of commits of worker \textit{i} \\
        $\Delta c_i$     & the number of commits betweentwo checks for worker \textit{i} \\
        $\bar{c}$ 	     & the average number of commits of P workers \\
        $\overline{\Delta c_i}$   & the average number of commits between two checks for \textit{m} workers \\  
        $\widehat{\Delta c_i}$    & the expect number of commits between current check and next check \\
        $\Delta t$ 		& the interval time between two checks \\
        $t_i$ & the interval time between two pushes for worker \textit{i} \\ 
      \bottomrule[1.5pt]
    \end{tabular*}
  \end{minipage}
\end{table*}

\subsection*{Appendix C}

Suppose parameter servers check the numbers of commit every $\Delta_t$ unit time.
\textit{B}: the bandwidth of PS;
\textit{S}: model size;
$t_i$: The interval between two submissions;
\textit{link}: \# of workers exchanging data with the PS during a certain period of time;
\textit{m}: mini-batch size.
\\
\begin{equation}
\begin{aligned}
 t_p & = (T_{preprocessing} + m \cdot T_{forward} + T_{back}) \cdot \Delta_p \\
 	&+ \frac{2S}{\frac{B}{link}} + T_{update}\cdot link + T_{overhead} \\
          & = \theta_0 \cdot \Delta_p  + \theta_1
\end{aligned}
\end{equation}
So, we can estimate the $\Delta c_p$:
\begin{equation}
\begin{aligned}
\Delta c_p = & \frac{\Delta_t}{t_p} = O(\frac{1}{\Delta_p}) = \\
                 = & \frac{1}{\alpha_0 \cdot \Delta_p + \alpha_1} + \alpha_2
\end{aligned}
\end{equation}
Simulate this formula.
$$\Rightarrow optimize: \sum_{p = 1}^{P}((c_{p} + \Delta c_p) - (\bar{c}+ \overline{\Delta c}_p))^2 $$
\cite{1998kukar_cost}
}

\section{More details on Evaluation}

\subsection{Application}
(1) {\em Image Classification.} We use the Cifar-10 \cite{cifar10} as the dataset, which consists of 6 files; we use the first 4 files (40,000 images) for training, the 5th file (10,000 images) for model evaluation during training and the last file (10,000 images) for testing the trained model (e.g., obtain model accuracy). 
A CNN model is trained on Cifar-10, which is from the TensorFlow tutorial \cite{tf_cifar10_tutorial}.
This CNN model can be trained within several hours with our 18 workers.

(2) {\em Fatigue life prediction.} 
As a practical edge-learning example, we predict the fatigue level of components on a high-speed train, using our dataset collected from the China high-speed rail system. Taking bogies on a train as an example, we train a recurrent neural network (RNN) model (5MB in size) to predict future fatigue levels of bogies on the trains, to decide whether each bogie needs to be repaired. The dataset includes the stress of bogies (collected by multiple sensors deployed on the trains), the temperature, component age, etc. Our input features for RNN training include: (i) historical stresses of the bogie; (ii) the age of a bogie; (iii) starting point and destination of the train where the bogie is located (as different trains running on different lines experience different weather, rail condition and continuous working hours); (iv) temperature, which significantly influences physical characteristics of metal. The RNN produces one of the three fatigue levels: \textit{0} representing that a bogie does not need to be repaired, \textit{1} denoting minor repair in need, and \textit{2} showing replacement is needed.

(3) {\em Coefficient of Performance (COP) prediction}
As another practical edge-learning example, we predict the COP of chillers, i.e., the ratio of refrigeration effect generated by a chiller against its energy consumption, which is a widely used metric for evaluating chiller performance \cite{mathur1988determining}. We exploit a chiller dataset, collected from 5 chillers each day during 2012 to 2015. Each record in the dataset consists of the outlet temperature produced by the chiller, the outdoor temperature, electricity consumption on that day, the age of the chiller and some other features that may influence the refrigeration effect. These records are labeled with the COP of that day. We train a global linear SVM model with electricity consumption and some other features as input, and produce COP as the output. 

\subsection{Experiment Results}

\subsubsection{System Scalability}\label{sec:scaling}
\begin{figure}[!th]
\begin{center}
	\includegraphics[width = .95\columnwidth]{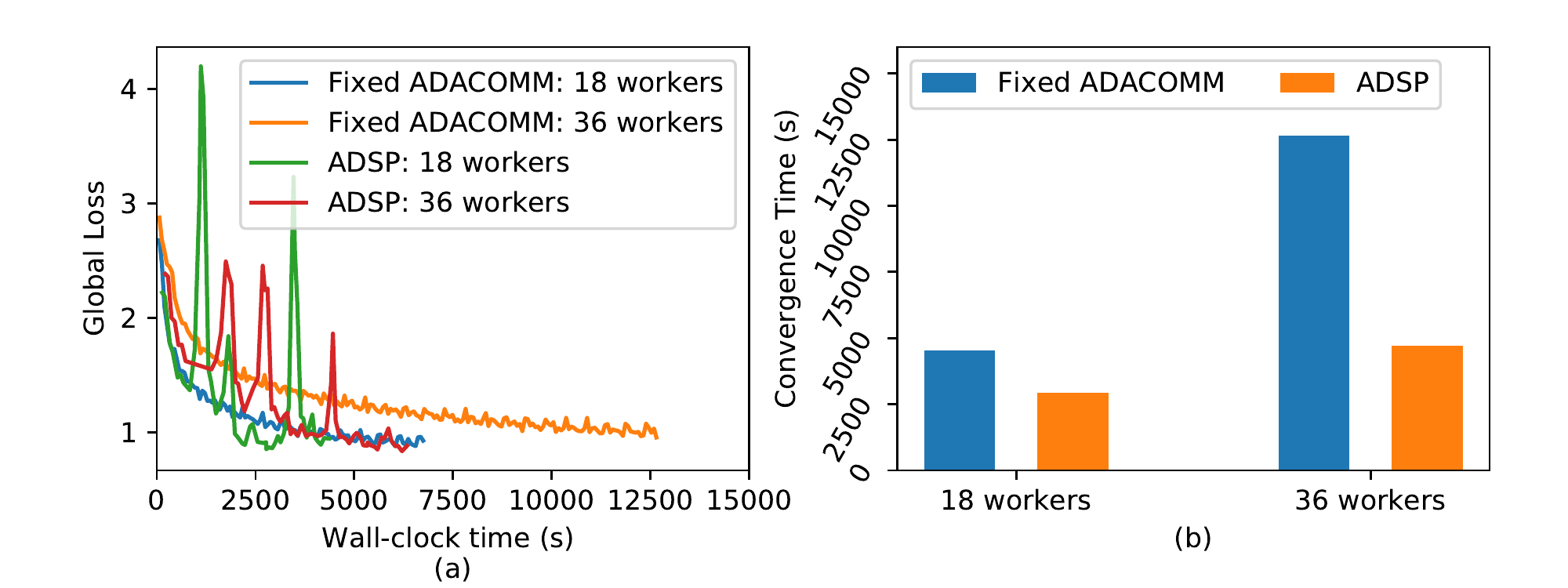}
	\caption{Comparison of ADSP with {\em Fixed ADACOMM} at different system scales}
	\label{figure:scale}
\end{center}
\end{figure}

We further evaluate ADSP with 36 workers used for model training, whose hardware configuration follows the same distribution as in the 18-worker case. Fig.~\ref{figure:scale} shows that when the worker number is larger, both ADACOMM and ADSP become slower, and ADSP still achieves convergence faster than Fixed ADACOMM (which is more obvious than in the case of smaller worker number). Intuitively, when the scale of the system becomes larger, the chances increase for workers to wait for slower ones to catch up, resulting in that more time being wasted with Fixed ADACOMM; ADSP can use this part of time to do more training, and is hence a more scalable solution in big ML training jobs.

\begin{figure}[!th]
\begin{center}
	\includegraphics[width = .95\columnwidth]{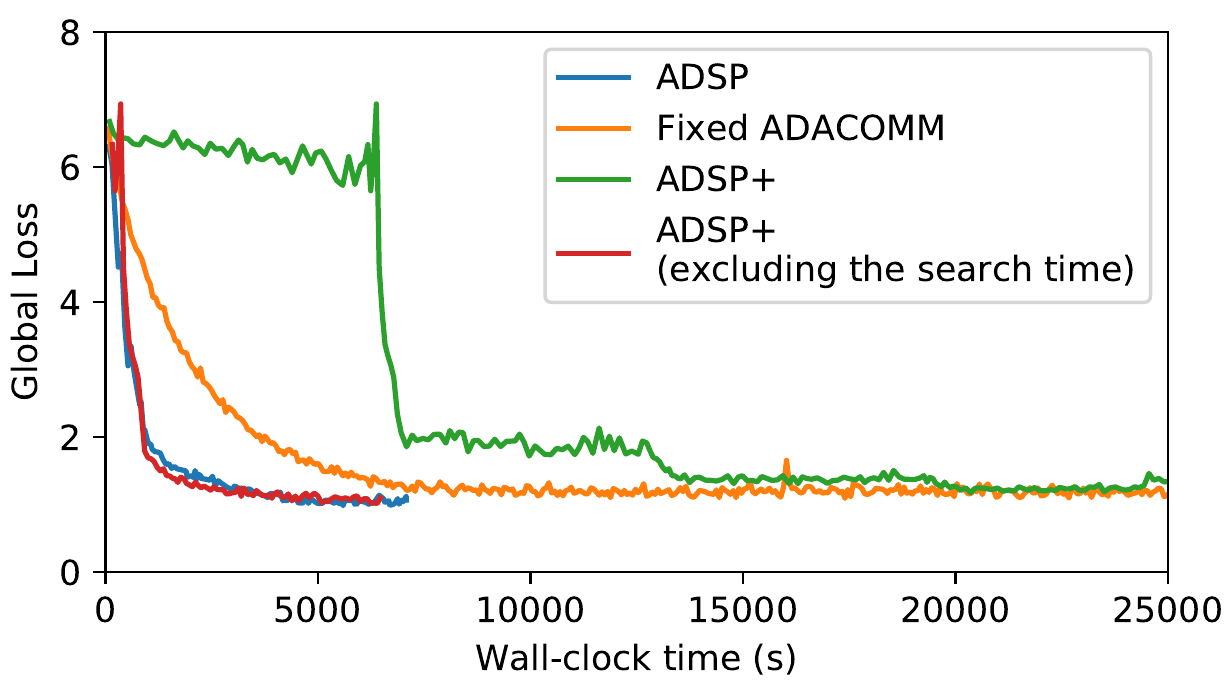}
	\caption{Verification of ADSP's near-optimality with non-waiting local updates}
	\label{figure:near-optimal}
\end{center}
\end{figure}

\subsubsection{The Impact of Waiting Time} 
\label{sec:nearoptimal}
We next look further into Fixed ADACOMM and ADSP. Both methods require workers to commit to the PS at the same commit rate; ADSP allows faster workers to accumulate as many local updates as their processing capacities allow between two commits; Fixed ADACOMM allows workers to train the same $\tau$ mini-batches between two commits, such that faster workers would still be blocked to wait for slower workers. A question arises that: what is the optimal number of extra mini-batches to train between two commits at a fast worker, beyond the same $\tau$ mini-batches, such that the global model convergence is fastest? In another word, does the maximal training approach in ADSP, i.e., no waiting at each worker, achieve the fastest convergence, or a slightly smaller number of local updates $\tau_i$ between two commits (but still larger than $\tau$) lead to better performance?

To answer the above questions, we implement a variant of ADSP, namely \textit{ADSP$^+$}, which searches all possible numbers of local updates to carry out at each worker between two commits to decide the best $\tau_i$'s in an offline fashion, given a fixed $C_{target}$. Obviously, \textit{ADSP$^+$} achieves the optimal convergence performance if we ignore the significantly large searching time (which is also the reason why we do not use it in implementing ADSP). We compare the performance of ADSP to \textit{ADSP$^+$} after deleting the search time of the latter. We evaluate each method on the Cifar-10 dataset with a large CNN model(around 14MB). Fig.~\ref{figure:near-optimal} shows that ADSP can achieve similar performance to \textit{ADSP Plus} (with $\tau_i$ search time excluded), which verifies that ADSP is near-optimal.

\subsubsection{The Impact of Batch Size}
\begin{figure}
\begin{center}
  \includegraphics[width = .95\columnwidth]{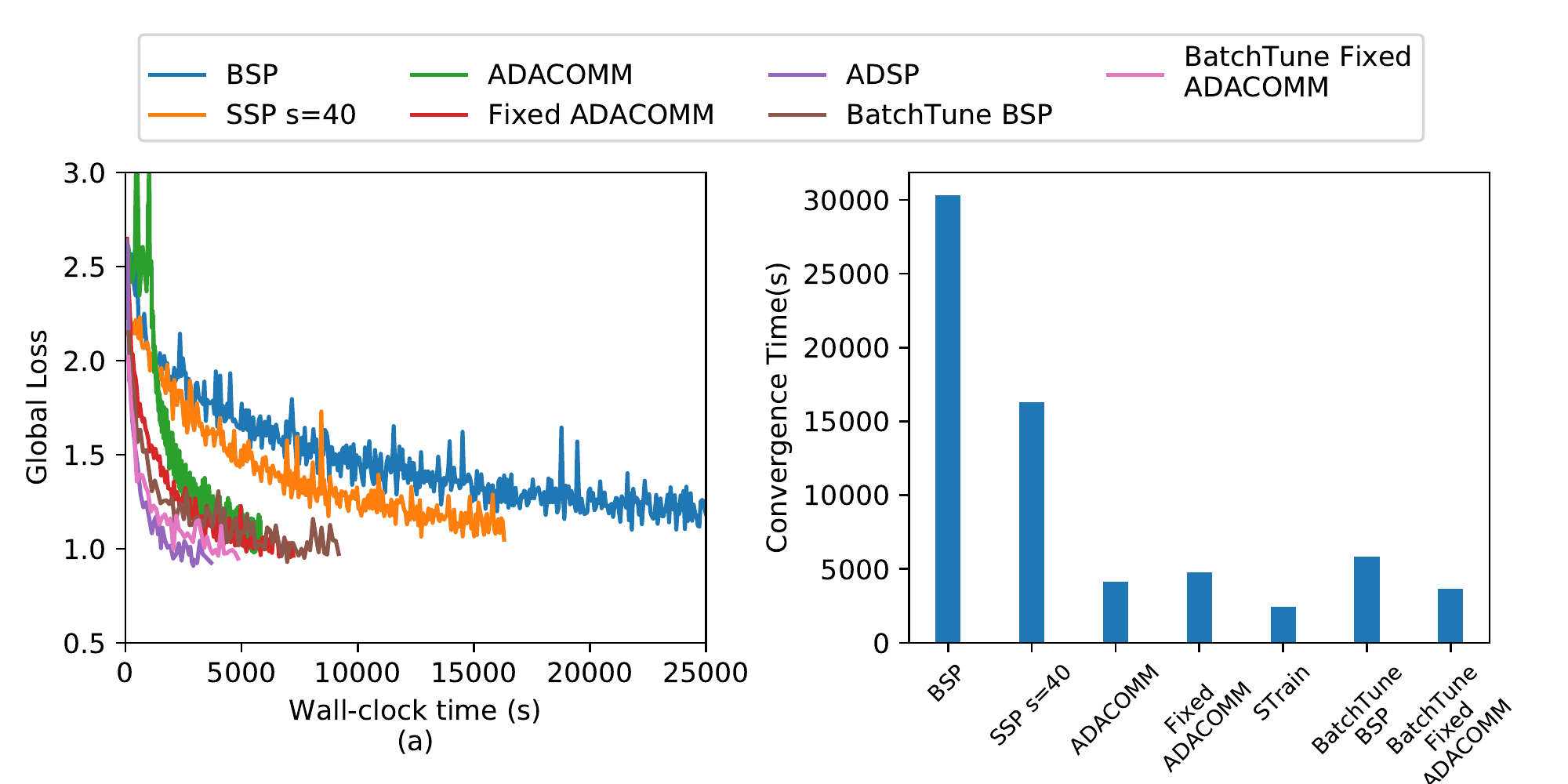}
  \caption{Comparison of ADSP with {\em BatchTune BSP} and {\em BatchTune Fixed ADACOMM}}
  \label{figure:r2sp}
\end{center}
\end{figure}

To mitigate resource wastage in a heterogeneous cluster, $R^2SP$ \cite{chenround} adaptively tunes the batch size to make full use of the computing capability, i.e., faster workers use a larger batch size while keeping the global batch size the same. We call this method of adapting to heterogeneity by tuning the batch size of each worker as {\em BatchTune}. To compare ADSP with {\em BatchTune}, we apply the method of tuning batch size to {\em BSP} and {\em Fixed ADACOMM}, neither of which are designed for heterogeneous scenarios, and call them {\em BatchTune BSP} and {\em BatchTune Fixed ADACOMM}, respectively. Fig.~\ref{figure:r2sp} shows that after applying {\em BatchTune}, both {\em BSP} and {\em Fixed ADACOMM} converge significantly faster, which means tuning batch size can indeed alleviate the influence of heterogeneity. Nonetheless, Fig.~\ref{figure:r2sp} further tells that ADSP still performs best in terms of the convergence speed. The reason is twofold: firstly, ADSP's online search algorithm can automatically find the optimal commit rate, thus more flexible, while {\em BatchTune} is restricted by the fixed global batch size; secondly, the difference between {\em BatchTune} and ADSP resides whether to use one larger batch or more small batches for the faster workers, and SGD with small batches is much faster \cite{2017goyal-accurate}. 

\subsubsection{The Impact of Bandwidth}
\begin{figure}
\begin{center}
  \includegraphics[width = .95\columnwidth]{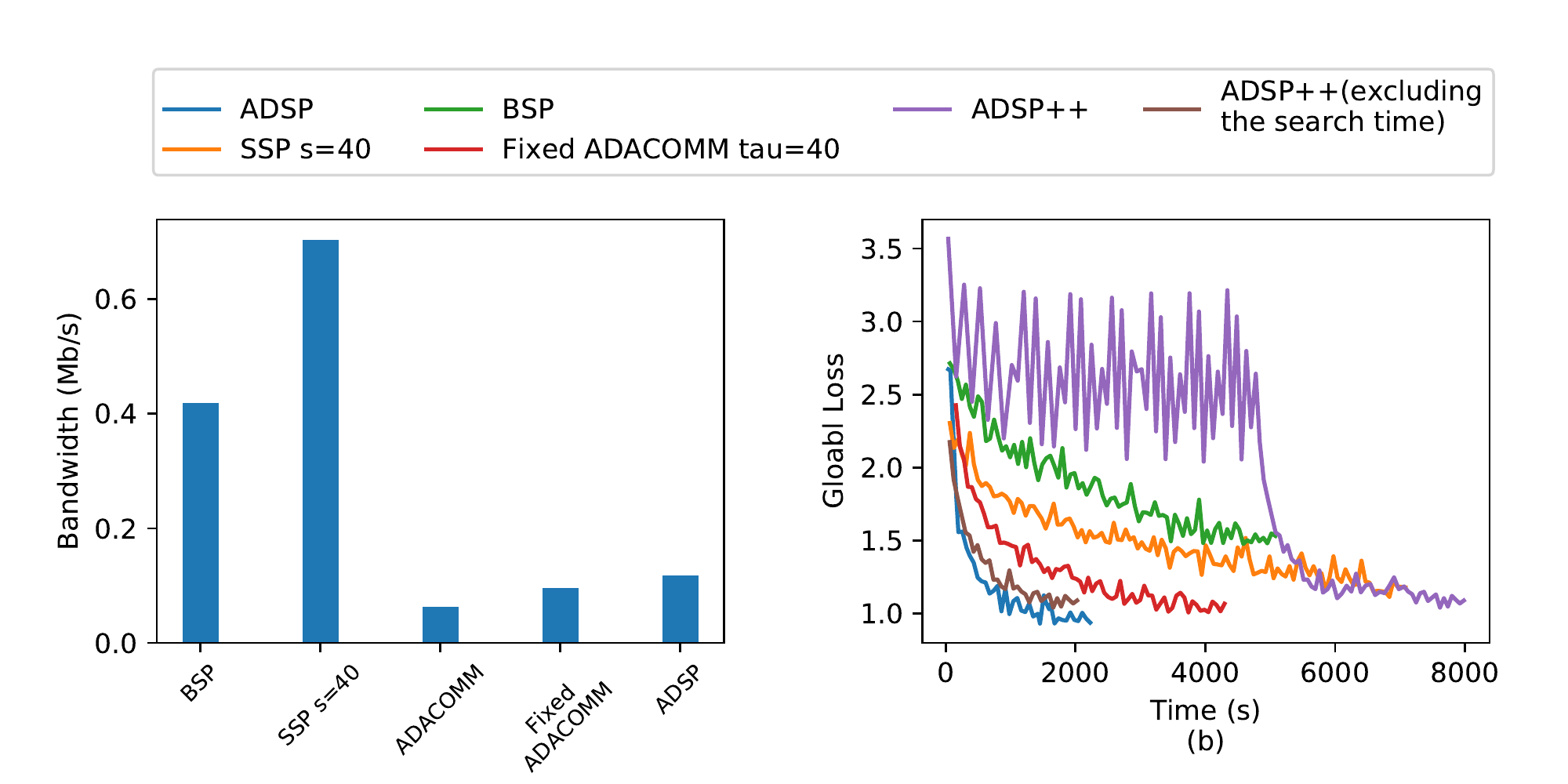}
  \caption{(a) Comparison of Bandwidth usage. (b) Comparison of ADSP with {\em ADSP$^{++}$}.}
  \label{figure:bandwidth}
\end{center}
\end{figure}

We compare the bandwidth usage of the systems based on the amount of data transmitted between workers and the parameter server per unit time. As shown in Fig.~\ref{figure:bandwidth}(a), the bandwidth usage of {\em BSP} and {\em SSP} is very large, since they both force each worker to commit updates to the parameter server in every step. The bandwidth usage of {\em ADSP}, although smaller than that of {\em BSP} and {\em SSP}, is slightly larger than that of {\em Adacomm} and {\em Fixed Adacomm}. The bandwidth usage of {\em ADSP}, {\em Adacomm} and {\em Fixed Adacomm} is determined by the the commit rate -- $\Delta C_{target}$ for {\em ADSP} and $\tau$ for both {\em Adacomm} and {\em Fixed Adacomm}. The larger the commit rate is, the more bandwidth it takes.
Since ADSP starts with a small commit rate $\Delta C_{target}$ and searches the optimal commit rate for the fastest convergence by increasing $\Delta C_{target}$ by 1 each time, taking bandwidth usage into consideration, the optimal commit rate searched by Alg.~\ref{algorithm:scheduler} should still be the best choice -- the optimal point which can reach the best convergence speed with as small bandwidth usage as possible.
\subsubsection{The Impact of Hyperparameters}
For machine learning, the convergence speed is highly related to some hyper-parameter settings, including learning rate, momentum, etc. Here we implement another version of {\em ADSP}, called {\em ADSP$^{++}$}, which blocks the training process and searches for the optimal learning rate and momentum at the beginning of each epoch. In Fig.~\ref{figure:bandwidth}(b), we compare {\em ADSP$^{++}$} with and without the search time with {\em ADSP}. We observe that ADSP achieves a comparable convergence speed even compared with {\em ADSP$^{++}$} after excluding the search time.

\subsubsection{The Impact of Model Size}
\begin{figure}
\begin{center}
  \includegraphics[width = .95\columnwidth]{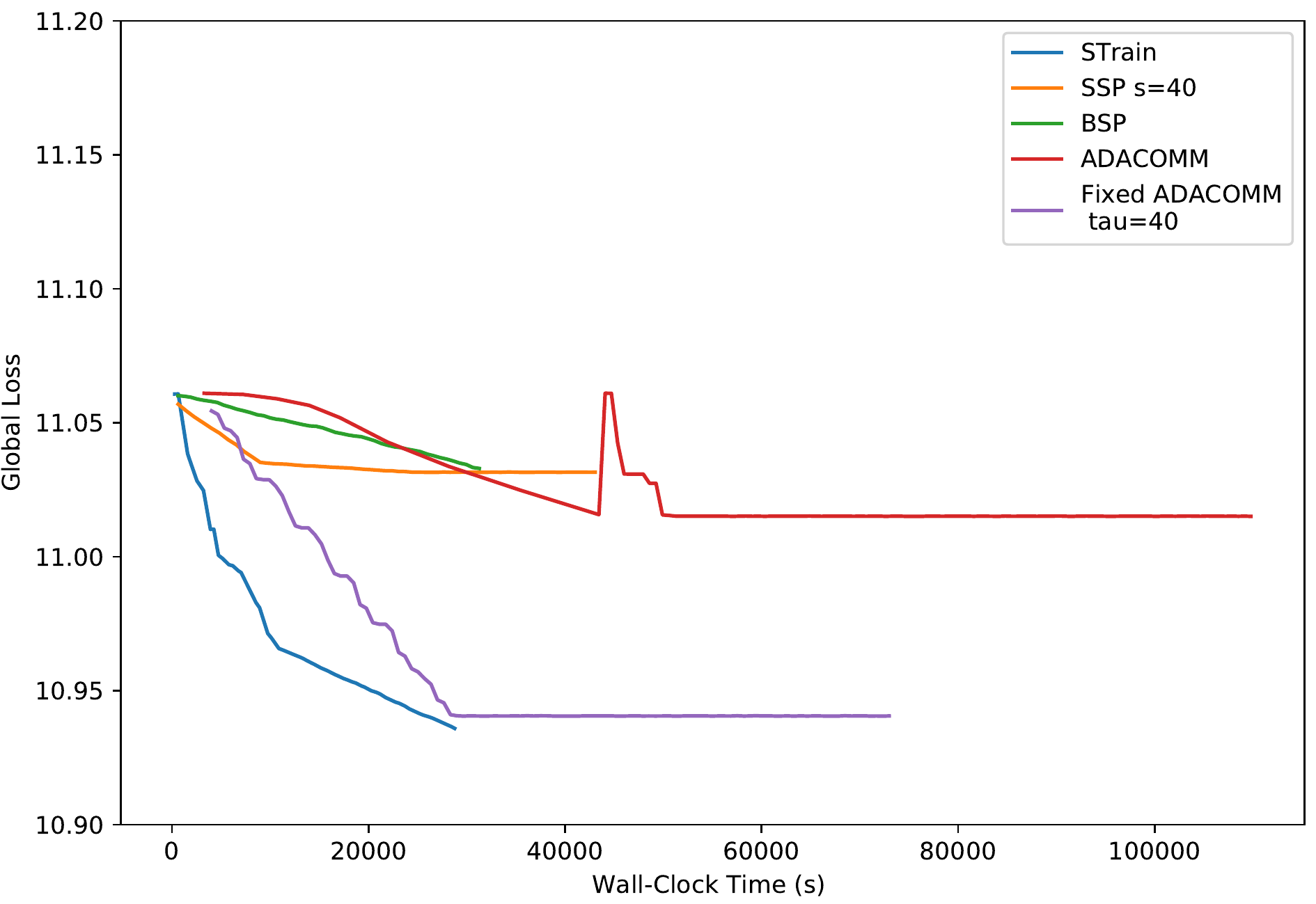}
  \caption{Comparison of ADSP with baselines running a large model}
  \label{figure:vgg}
\end{center}
\end{figure}

To observe how ADSP performs with a large model, we evaluate using VGG-16 \cite{simonyan2014very}, a state-of-the-art convolutional neural network model for imageNet with a model size of 528 MB, and compare the convergence speed. Considering that the time to process a batch of data increases a lot, we reduce the batch size to 32 and increase the check period $\Gamma$ to 600 seconds. Fig.~\ref{figure:vgg} shows that ADSP still outperforms the baseline systems even with an extremely large model. The reason is that it requires more processing time to train a batch with a large model and the waiting time induced by the heterogeneity becomes more significant. Benefited from the no-waiting strategy, ADSP can reduce the waiting time and shows fast convergence speed.
\begin{figure}[!t]
\begin{center}
	\includegraphics[width = .95\columnwidth]{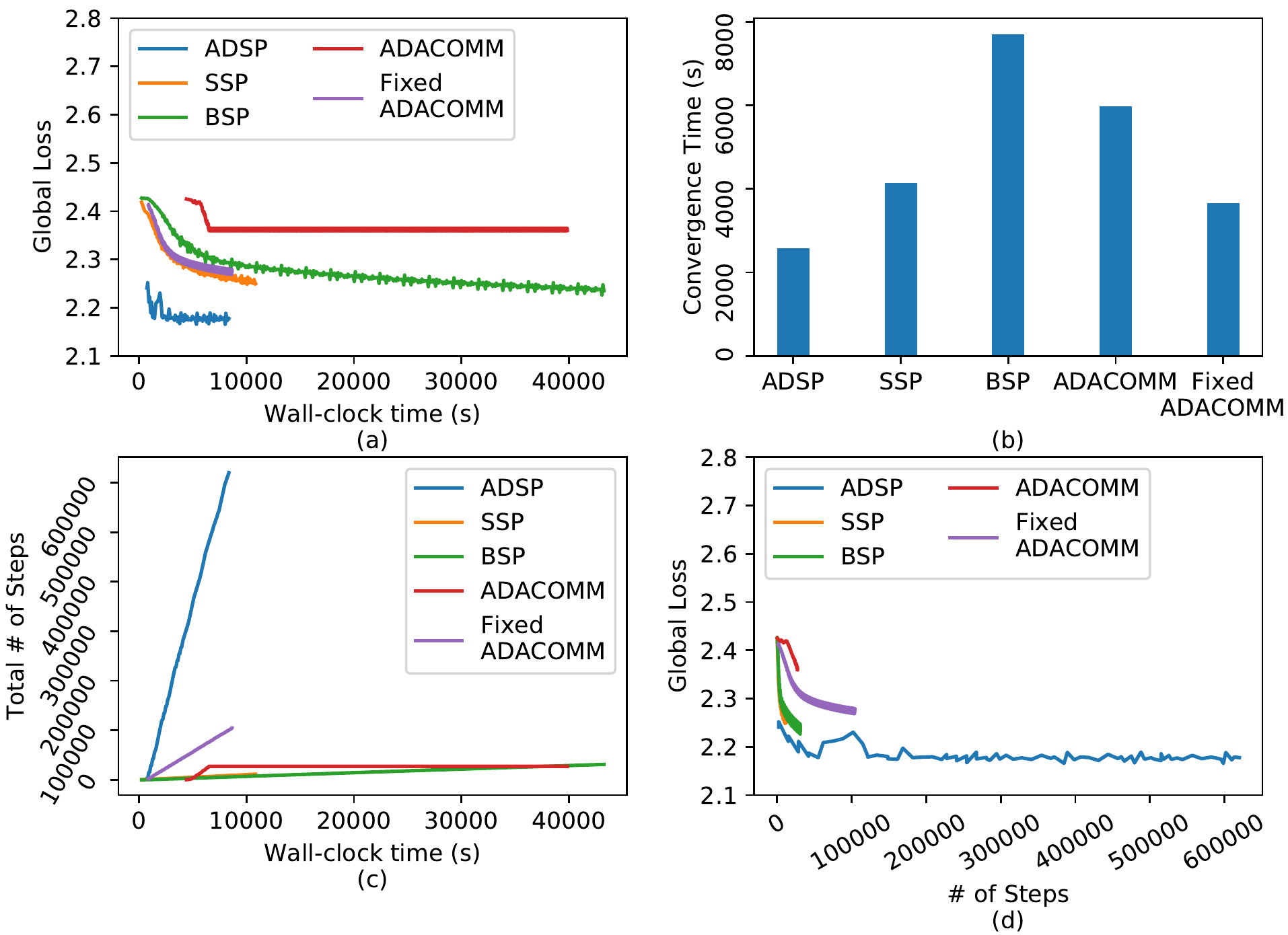}
	\caption{Comparison of ADSP with baselines 
	: training RNN on the high-speed rail dataset}
	\label{figure:rail_eval}
\end{center}
\end{figure}
\begin{figure}[!t]
\begin{center}
	\includegraphics[width = .95\columnwidth]{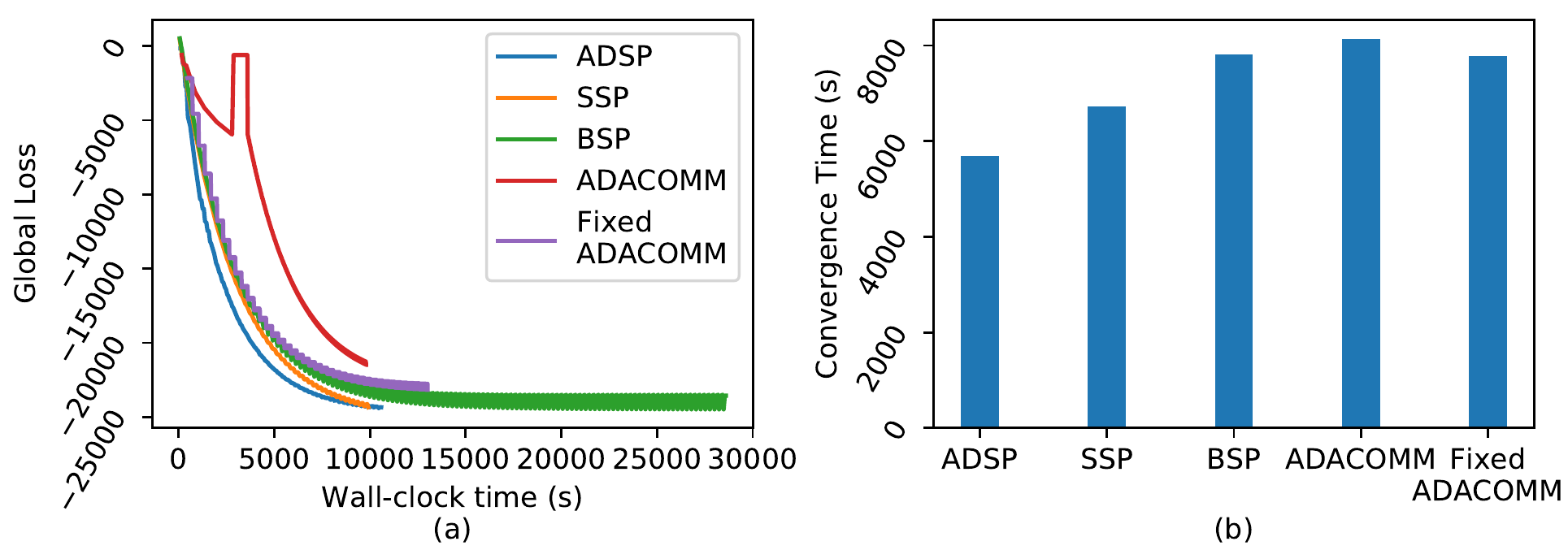}
	\caption{Comparison of ADSP with baselines 
	: training SVM on the chiller dataset}
	\label{figure:chiller_eval}
\end{center}
\end{figure}

\subsubsection{Performance on Fatigue-M and COP-M}
We further evaluate ADSP by training the RNN on the high-speed rail dataset and the linear SVM on the chiller dataset, respectively. 
Fig.~\ref{figure:rail_eval} shows evaluation results in the high-speed rail case. Similar to our previous observation, 
ADSP performs the best, and is faster than Fixed ADACOMM by $29.5\%$ in convergence time.
Similar observations can be made based on Fig.~\ref{figure:chiller_eval} for the chiller case.
\end{appendices}
\end{document}